\documentstyle[12pt,aaspp4,psfig]{article}

\newcommand{\km}{km s$^{-1}$}
\newcommand{\vy}[2]{#1_{\scriptscriptstyle #2}}
\newcommand{\Ly}{Ly$\alpha$}

\def\gtorder{\mathrel{\raise.3ex\hbox{$>$}\mkern-14mu
             \lower0.6ex\hbox{$\sim$}}}
\def\ltorder{\mathrel{\raise.3ex\hbox{$<$}\mkern-14mu
             \lower0.6ex\hbox{$\sim$}}}
\def\proptwid{\mathrel{\raise.3ex\hbox{$\propto$}\mkern-14mu
             \lower0.6ex\hbox{$\sim$}}}
\def\arcsec{\ifmmode '' \else $''$\fi}

\def\arcsecpoint{\ifmmode ''\!. \else $''\!.$\fi}

\def\kms{\ifmmode {\rm km\ s}^{-1} \else km s$^{-1}$\fi}
\def\Msun{\ifmmode {\rm M}_{\odot} \else M$_{\odot}$\fi}
\def\Lsun{\ifmmode {\rm L}_{\odot} \else L$_{\odot}$\fi}
\def\Zsun{\ifmmode {\rm Z}_{\odot} \else Z$_{\odot}$\fi}

\def\ergscm2{ergs\,s$^{-1}$\,cm$^{-2}$}
\def\qo{\ifmmode q_{\rm o} \else $q_{\rm o}$\fi}
\def\Ho{\ifmmode H_{\rm o} \else $H_{\rm o}$\fi}
\def\ho{\ifmmode h_{\rm o} \else $h_{\rm o}$\fi}

\def\vFWHM{\ifmmode v_{\mbox{\tiny FWHM}} \else
            $v_{\mbox{\tiny FWHM}}$\fi}
\def\CCF{\ifmmode F_{\it CCF} \else $F_{\it CCF}$\fi}
\def\ACF{\ifmmode F_{\it ACF} \else $F_{\it ACF}$\fi}
\def\Halpha{\ifmmode {\rm H}\alpha \else H$\alpha$\fi}
\def\Hbeta{\ifmmode {\rm H}\beta \else H$\beta$\fi}
\def\Hgamma{\ifmmode {\rm H}\gamma \else H$\gamma$\fi}
\def\Hdelta{\ifmmode {\rm H}\delta \else H$\delta$\fi}
\def\Lya{\ifmmode {\rm Ly}\alpha \else Ly$\alpha$\fi}
\def\Lyb{\ifmmode {\rm Ly}\beta \else Ly$\beta$\fi}
\def\Lyg{\ifmmode {\rm Ly}\beta \else Ly$\gamma$\fi}

\def\ciii{\ifmmode {\rm C}\,{\sc iii} \else C\,{\sc iii}\fi}
\def\civ{\ifmmode {\rm C}\,{\sc iv} \else C\,{\sc iv}\fi}

\def\o5007{[O\,{\sc iii}]\,$\lambda5007$}

\def\o{\o}
\begin{document}
\title{HST OBSERVATIONS OF THE BROAD ABSORPTION LINE QUASAR PG 0946+301}

\author{Nahum Arav\footnote{IGPP
                LLNL, L-413,
                P.O. Box 808,
                Livermore, CA 94550 \\  I: narav@igpp.llnl.gov},
Kirk T. Korista\footnote{Western Michigan Univ.,
Dept. of Physics,
1120 Everett Tower,
Kalamazoo, MI  49008}$^,$\footnote{
University of Kentucky
Dept. of Physics \& Astronomy
Lexington, KY  40506},
Martijn de Kool\footnote{JILA,  University of Colorado 
Boulder, CO  80309-044}$^,$\footnote{Astrophysical Theory Centre, 
John Dedman Building, ANU ACT
0200, Australia}, 
Vesa T. Junkkarinen\footnote{Center for Astrophysics and Space Sciences,
UCSD, 9500 Gilman Dr. La Jolla CA 92093} 
\& Mitchell C. Begelman\footnote{JILA,  University of Colorado 
Boulder, CO  80309-0440}$^,$\footnote{Also at Department of Astrophysics, 
 University of Colorado, Boulder, CO  80309-0440}  }



\begin{abstract}

We analyze HST and ground based spectra of the brightest BALQSO in the
UV: PG~0946+301.  A detailed study of the absorption troughs as a
function of velocity is presented, facilitated by the use of a new
algorithm to solve for the optical depth as a function of velocity for
multiplet lines.  We find convincing evidence for saturation in parts
of the troughs.  This supports our previous assertion that saturation
is common in BALs and therefore cast doubts on claims for very high
metallicity in BAL flows.  Due to the importance of BAL saturation we
also discuss its evidence in other objects.  In PG~0946+301 large
differences in ionization as a function of velocity are detected and
our findings supports the hypothesis that the line of sight intersects
a number of flow components that combine to give the appearance of the
whole trough.  Based on the optical depth profiles, we develop a
geometrical-kinematical model for the flow.  We have positively
identified 16 ions of 8 elements (H~I, C~III, C~IV, N~III, N~IV, N~V,
O~III, O~IV, O~V, O~VI, Ne~V, Ne~VIII, P~V, Si~IV, S~V, S~VI) and have
a probable identifications of Mg~X and S~IV.  Unlike earlier analysis
of IUE data, we find no evidence for BALs arising from excited ionic
states in the HST spectrum of PG~0946+301.


\end{abstract}

\section{INTRODUCTION}

Broad absorption troughs associated with prominent resonance lines
such as C IV $\lambda$1549, Si IV $\lambda$1397, N V $\lambda$1240,
and \Ly\ $\lambda$1215 appear in about 10\% of all radio quiet quasars
(Foltz et al. 1990).  These troughs are known as broad absorption
lines (BALs) and are commonly attributed to material flowing out from
the vicinity of the central engine.  Typical velocity widths of the
BALs are $\sim10,000$ \km\ (Weymann, Turnshek, \& Christiansen 1985;
Turnshek 1988) with terminal velocities of up to 60,000 \km.  These
high terminal velocities are seen in H1414+089 (66,000 \km, Foltz et
al. 1983) and in Q1231+1320 (58,000 \km, spectrum from Korista et
al. 1993). The percentage of BALQSOs among all quasars combined with
the assertion that the flow covers $\ltorder 0.2$ of the sky on
average as viewed from the nucleus of the QSO (Hamann, Korista, \&
Morris 1993), suggests that all QSOs have BAL flows. This conclusion
is supported by the similarity of the broad emission lines (BELs) in
the spectra of BALQSOs and non-BALQSOs (Weymann et al. 1991).  The BAL
region is probably situated further away from the continuum source
than the broad emission line region based on the attenuation of the
BELs by the BALs (Junkkarinen, Burbidge, \& Smith 1983; Turnshek 1988;
for alternative view see Murray et al. 1995). 
Radiative acceleration probably governs the dynamics of the flow
(Arav, Li \& Begelman 1994; Murray et al. 1995; de Kool \& Begelmen 1995;
Arav et al. 1995)

Establishing the physical properties of
the flow by determining the ionization equilibrium and
abundances (IEA) of the BAL material is a fundamental issue in 
BALQSOs studies.  Data from HST are crucial in
trying to answer these questions. Ground based observations show only
a small number of BALs (most of which are listed above) that arise
from different elements.  HST observations of moderate to high
redshift ($z=1-3$) objects allow us to access the 500--1000 \AA\
rest-wavelength region which contains many more BALs, including BALs
from different ions of the same element, and even different BALs from
the same ions.    Without data on BALs from
different ions of the same element the effects of ionization and
abundances are very difficult to decouple. The first and heretofore the
most comprehensive study of a BALQSO using HST data was done by
Korista et al. (1992, hereafter K92) on BALQSO 0226--1024.  In the
combined HST and ground based spectrum at least 12 different absorbing
ions were identified including the following CNO ions: C III, C IV, N
III, N IV, N V, O III, O IV, O VI.  Several groups (Korista et
al. 1996; Turnshek et al. 1996; Hamann 1996) have used these data in
their IEA studies while introducing innovative theoretical approaches
to the problem.
      
However, these works suffer from large uncertainties, which comes from
reliance on {\it apparent} integrated ionic column densities
($N_{ion}$).  Accurate $N_{ion}$ are crucial because inferences on the
IEA in the BAL region are derived by trying to simulate BAL column
densities using photoionization codes.  The problem is that the
measured apparent optical depths in the BALs (defined as $\tau=-ln(I_r)$, where
$I_r$ is the residual intensity seen in the trough) cannot be directly
translated to realistic $N_{ion}$ due to loose constraints on the
covering factor and level of saturation (see discussion in K92).  We
note that the term ``covering factor'' is defined in this paper as the
portion of the radiating source covered by the flow as seen by the
observer.  This is to be distinguished from the percentage of solid
angle which the flow covers as viewed fron the nucleus of the QSO.
Arav (1997) has demonstrated that in the case of BALQSO 0226--1024 the
major BALs are very probably saturated although not black.  If the
BALs are saturated, the inferred $N_{ion}$ are only lower limits, and
thus the previous conclusions regarding the IEA in this object, and by
extension in all BALQSOs, are very uncertain. By now the body of
evidence for non-black BAL saturation has grown considerably and in
\S~4.2 we discuss these in detail.

Therefore, a crucial step in BAL IEA studies is a careful analysis of
the BAL profiles aimed at determining the real optical depths and thus
real column densities.  In order to do so we have to abandon the use
of integrated column densities.  Different parts in the BALs can (and
do, see below) show different levels of saturation and covering
factors, and the apparent optical depths show velocity-dependent
ionization changes.  Supporting evidence comes from the so called
mini-BALs, which are defined as intrinsic absorbers with velocity
width smaller than 2000 \km, whereas traditional BALs are wider then
2000 \km (Weymann et al. 1991).  Mini-BALs show large variation in
covering factor on scales of hundreds of \km\ (Barlow 1997).  Thus, in
order to shed light on the IEA of BAL flows, there is no substitute
for a detailed study of optical depth as a function of velocity
($\tau(v)$) for as many BALs as possible in a given object.

 \begin{figure}[htb]
\centerline{\psfig{file=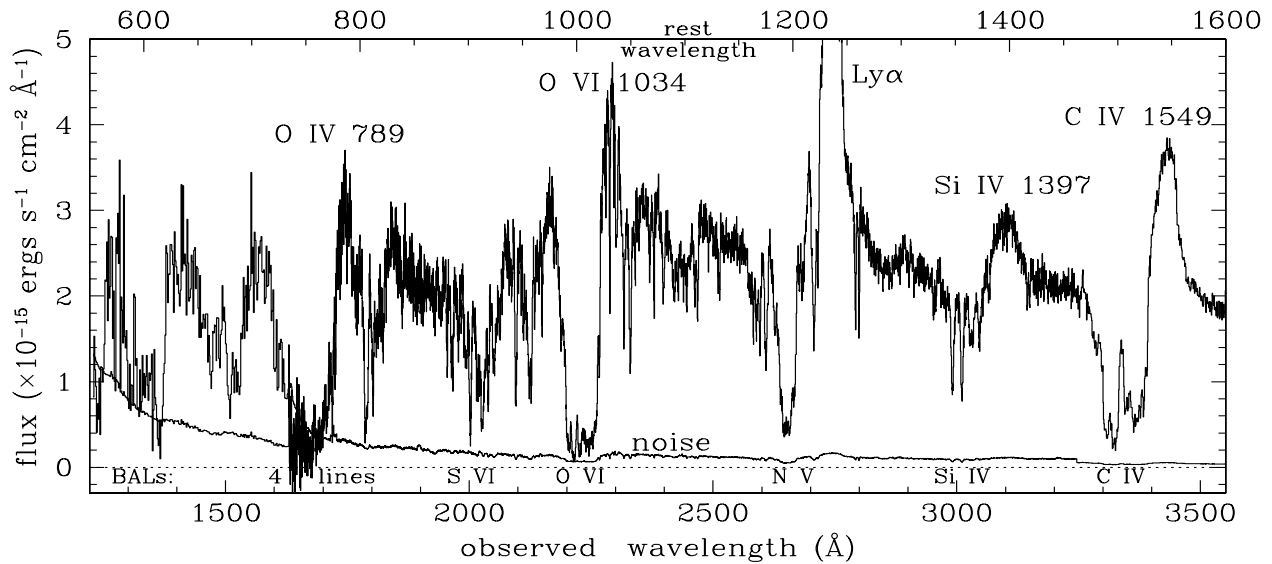,angle=-90,height=12.cm,width=18.cm}}

\caption{Composite spectrum of PG~0946+301, flux is measured in the observed frame.}\label{fig-1}
\end{figure}

To this effect, we analyze the spectrum of PG 0946+301, which was
discovered by Wilkes (1985) and observed in the UV by Pettini and
Boksenberg (1986) using the IUE satellite.  This object is by far the
most suitable for such an analysis.  From all the BALQSOs that were
observed with HST it is the brightest with observable lines down to
570 \AA\ rest frame, and has at least five times higher UV flux,
shortward of 1000 \AA\ rest frame, than any observed object.  Unlike
the cases of BALQSO 0226--1024 and a few other BALQSOs in the HST
archive, the flux of PG 0946+301 does not diminish rapidly shortward
of 1000 \AA.  Another important advantage is that due to its low
redshift (z=1.223), the \Ly\ forest in the spectrum of PG 0946+301 is
much less dense than in other observed BALQSOs and therefore the BALs
are less contaminated with unrelated absorption.  These advantages are
crucial since in order to make a detailed $\tau(v)$ analysis much
higher quality ``clean'' data are needed than for the apparent-$N_{ion}$
approach. This analysis is based on a new algorithm we developed that
enables us to solve for the optical depth of the line whether it is a
doublet, triplet or higher order multiplet.  Solving for multiplets
other than doublets was not possible with the algorithm published by
Junkkarinen et al. (1983; which was also used in a modified way by
K92). Our algorithm is also more stable and can be used to extract
apparent-optical depth under saturated conditions.  Grillmair \&
Turnshek (1987) used a different algorithm, but give too few details
to allow a comparison.  We give a detailed description of our optical
depth solution method in Appendix A.

Supplementing the $\tau(v)$ analysis we also present results from the
traditional apparent-$N_{ion}$ approach.  This is done for two reasons. First,
there are  regions in the spectrum where several BALs are blended
together and therefore a $\tau(v)$ analysis in these region is
impossible.  In these regions the $N_{ion}$ analysis allow us to
obtain reasonable estimates for the existence and importance of
different BALs.  Second, the quality of available data is much lower
shortward of 750~\AA\ (rest frame).  and does not allow a detailed $\tau(v)$
analysis.  Several important lines reside in that part of the data and
our only inferences about them can be obtained by the 
apparent-$N_{ion}$ analysis.

The plan of the paper is as follows: In \S~2 we describe the data
acquisition and reduction.  In \S~3 we analyze the apparent optical
depth of the BALs.  Evidence for non-black saturation in PG 0946+301
is presented in \S~4 supplemented by a summery of evidence from other
objects.  Results from the integrated $N_{ion}$ approach (template
fitting procedure) are given in \S~5. A kinematic model for the flow
is presented in \S~6. In \S~7 we discuss the significance of
these findings and address the meta-stable lines issue.  Appendix A describes
the methods to extract $\tau(v)$ for different lines, and Appendix B
identifies the non-BAL absorption seen in the spectrum.

\section{DATA ACQUISITION AND REDUCTION}

HST FOS observations of PG 0946+301 were made on 16 February 1992 as
part of the FOS GTO program.  A low resolution spectrum was obtained
with the FOS Blue side using the rapid mode, the G160L grating, and
the 1.0{\tt"} aperture.  The integration time with the G160L was 1768
seconds.  The G160L spectrum has a FWHM resolution of about 8.0 \AA \
and a sampling of about 1.74~\AA \ per pixel.  Observations were
obtained with the FOS Red side on the same date with both the G190H
and G270H gratings with the 1.0{\tt"} aperture and exposure times of
4441 and 2663 seconds respectively.  The G190H resolution is 1.5~\AA \
FWHM with 0.36 \AA \ per pixel sampling and the G270H resolution is
2.2~\AA \ FWHM with 0.51 \AA \ per pixel sampling.  The usual reduction
procedure was used to generate flux as a function of wavelength.  The
G160L Blue and G190H Red observations were corrected for particle
background and any wavelength independent scattering using spectral
regions where the sensitivity is essentially zero.  The data were
corrected for detector and spectrograph sensitivity variations using
flat field observations close in time to the actual observations.
Since the flat fields of the FOS vary in time, the flat field
corrections sometimes leave residual features.  In these observations,
there are no apparent spurious features due to flat field variations.
Spectra from the three HST gratings were combined to obtain a full HST
spectrum. The relative wavelength determination was done by using
common features at the edges of the gratings. We estimate the relative
wavelength error to be less than half a pixel width.  
The absolute wavelength calibration was determined by the Galactic Mg~II
absorption lines seen in the G270H grating, assuming they lie at their
vacuum wavlengths in the frame of our Galaxy.

Optical observations of PG 0946+301 were made at Lick Observatory on
28 March 1992 (UT) using the Kast double spectrograph system on the
Shane 3m telescope.  A 3000s observation was obtained using the blue
side of the spectrograph with a 600 line mm$^{-1}$ grism blazed at
4310~\AA.  The spectrum has a useful wavelength range of 3150 -- 5250
~\AA \ with a resolution of about 5 \AA \ FWHM.  The usual optical
reduction procedures were followed using the VISTA reduction package.

\section{ANALYSIS OF THE BALs APPARENT $\tau(v)$}
In Figures 2 we present $\tau(v)$ solutions for seven BALs.  A
detailed description of the method used to extract these solutions
from the data is given in appendix A.  In this section we describe and
analyze the apparent optical depth solutions.  The apparent $\tau(v)$
is extracted under the assumptions that the source is completely
covered, and that the bottoms of the troughs are not filled-in by
scattering or some other light source.  We identify two major
components (A and B) and three minor components (C, D and E) in the
flow.  In figure 2 the $\tau(v)$ are organized according to the
ionization energy needed to destroy the ion, with the highest
ionization stage at the top.

 \begin{figure}
\centerline{\psfig{file=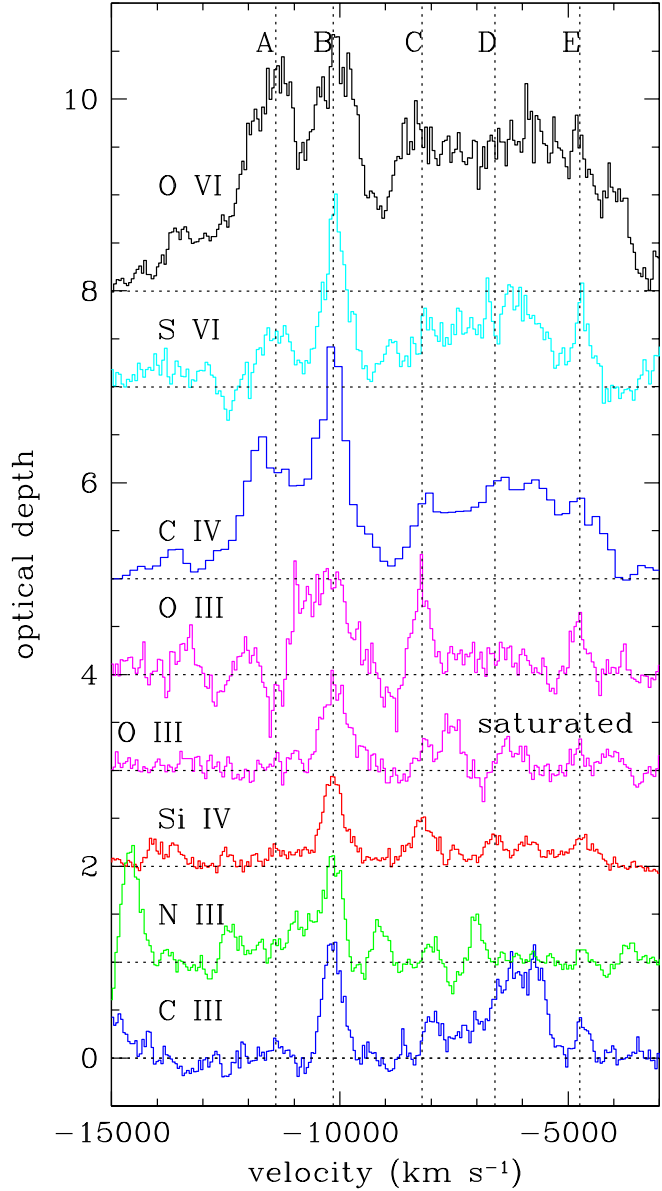,height=18.0cm,width=14cm}}
\caption{$\tau(v)$ for seven lines.  The horizontal dotted line at the
base of each $\tau(v)$ mark the $\tau=0$ for this line.  Vertical
dotted lines identify the components discussed in the text.  Both an
optically thin and a saturated solution of O~III are shown (see
\S~4.1). For presentation purposes each solution is shown with a
$\Delta\tau=1$ offset with respect to the one beneath it. The
exception is S~VI which for clarity has an offset of $\Delta\tau=2$
from the C~IV solution. }\label{fig-2}
\end{figure}

\subsection{The Different Components}
Component A is clearly detected only in the high ionization lines:
O~VI, S~VI and C~IV. It is strongest in O~VI ($\tau_{max}=2.3$), 
where its width is 1200
\km.   In S~VI component A is very similar in shape and
position to the O~VI one with $\tau_{max}=0.6$. Across most of
component A profile $\vy{\tau}{O~VI}/\vy{\tau}{S~VI}\simeq constant$,
this is in contrast to the situation in Component B (see Fig. 3 and
discussion below).  A small difference is seen at high velocity where
the O~VI seems to be somewhat more extended.  In C~IV component A is
also strong ($\tau_{max}=1.3$), but the location of the peak is
shifted by $\sim$400 \km\ to the blue with respect to those of O~VI
and S~VI.  However, the C~IV data has lower resolution than the HST data
and the position of the peak is based on only two pixels.
  C~III and Si~IV show marginal optical depth at the velocity
position of component A and the other low ionization lines have
$\tau(\mbox{A})\simeq 0$.

Component B is the most prominent feature in each BAL.  
The width of
component B differs substantially in different lines. In the high
ionization lines O~VI and C~IV the full-width-half-maximun (FWHM) of
component B is 1100 and 800 \km\ respectively,   whereas in the low
ionization lines the FWHM is $\simeq$500 \km\ with very similar component
profile.  However, the S~VI line does not adhere to the separation based
on ionization scheme.  Although S~VI ionization potential is higher
than that of C~IV, its component B profile is very similar to the
profiles seen in the low ionization lines (see Fig. 3).
 \begin{figure}
\centerline{\psfig{file=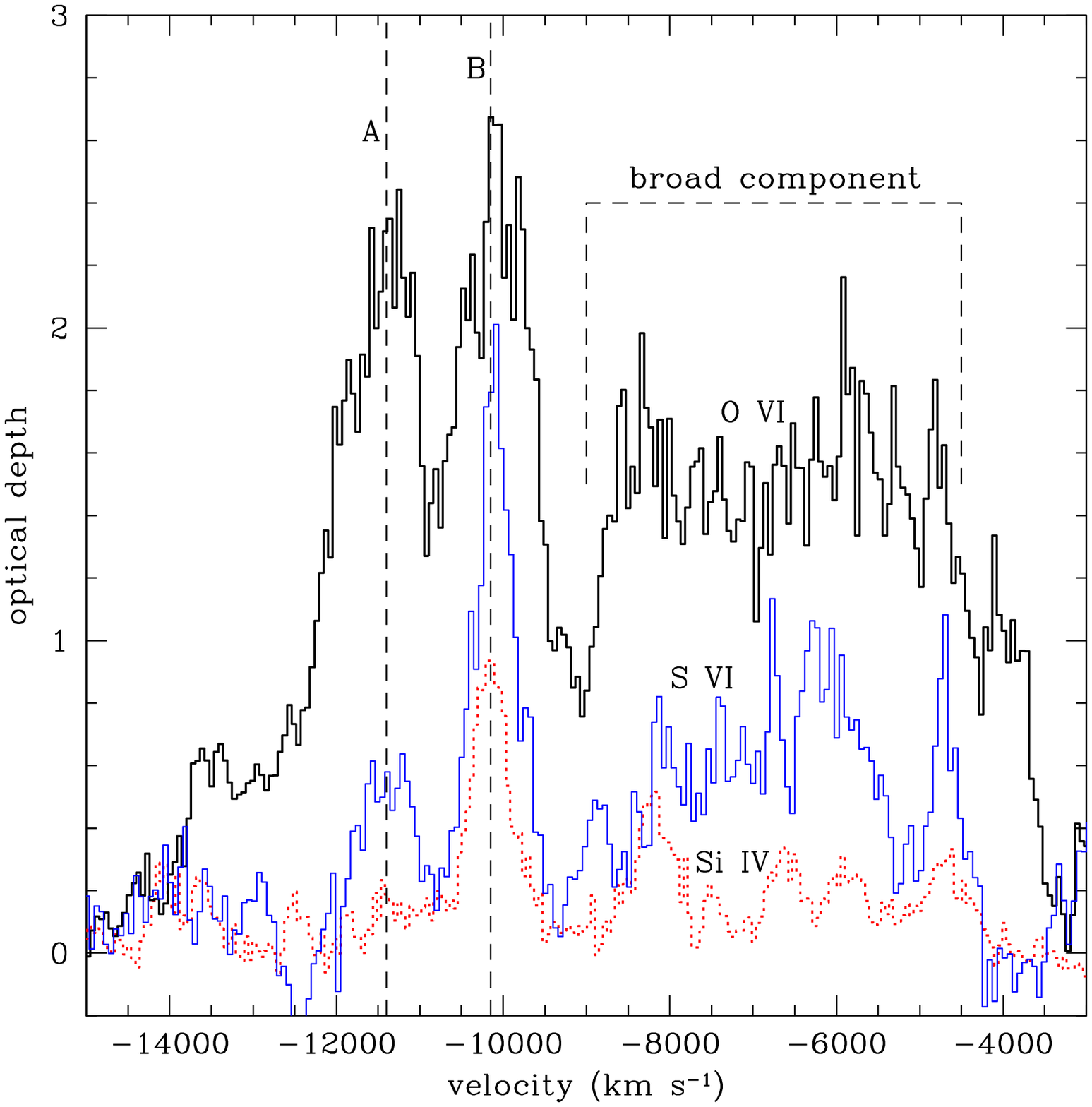,height=14.0cm,width=14cm}}

\caption{$\tau(v)$ for O~VI, S~VI and Si~IV, where the main components of
the flow are marked. }\label{fig 3}
\end{figure}

Being a singlet, the C~III line provides the most reliable profile for
component B.   It is reassuring that the profiles
of component B in Si~IV, saturated O~III (see \S~4.1) and S~VI are essentially
identical with that of C~III (within the quality of the data and the
uncertainties about the continuum level). The excess optical depth in C~III
near 6000~\km\/ is due to the N~III BAL. Component B in N~III
shows a deviation from this profile on its blue wing, which
is probably due to blending with component D of C~III. 

Component C 
does not stand out above the smooth, broad absorption in O~VI,  S~VI and C~IV.
It is strong in  Si~IV, and may possibly be associated with weak features in N~III,
C~III and saturated O~III with a small shift
($\sim200$ \km) to the red. The strong feature in the unsaturated O~III profile 
is most likely due to the deconvolution procedure, and shows clearly
the major effects that deconvolution with  inapropriate
 line ratios can introduce. 
 Component D is only seen in C~IV and Si~IV, and may be spurious.  
Component E is clearly seen in all seven BALs. The relatively
strong E component in S~VI might be partially due to component B of
P~IV~$\lambda$951, but the evidence is not strong enough for a clear
identification of this BAL.  Components C, D and E are significantly
narrower than components A and B.

\subsection{Ionization as function of velocity}
 We find strong
ionization changes as a function of velocity in the {\it apparent}
optical depth of different lines.  The most striking difference is
seen in component A, which is strong in the high ionization lines:
O~VI, C~IV, S~VI, but is very weak or 
absent from the low ionization lines Si~IV,
N~III and C~III. A lower limit for the optical depth ratio in
component A is $\vy{\tau}{O~VI}/\vy{\tau}{Si~IV}\gtorder 10$.  This is
in contrast with component B which appears in all lines and is only
twice as strong in the high ionization lines as in the low ionization
lines.  We conclude that in PG~0946+301 the apparent ionization has a
strong dependence on velocity. However, 
the actual physical situation is more complicated since 
 component B is saturated in all lines (see \S~4.1).
It is therefore possible that the ratios of real optical depths
show less dependence on velocity.  

A better diagnostic for the behavior of the real optical depth ratio
is available from the broad component between --9000 \km\ and --4500
\km\ (see Fig.~3), which encompasses components C, D and E.  This
component is much more prominent in the high ionization lines.  It is
deepest in O~VI with roughly a constant $\tau=1.5$.  The high,
flat-topped profile suggests that the line is saturated, and that the
remaining flux at the bottom of the absorption trough comes either
from a part of the continuum  source that is not completely
covered (partial covering) or consists of scattered light.  For
brevity, we refer to this combination of effects as ``saturation''. In
C~IV and S~VI the optical depth of the broad component is not
constant, but is still substantial ($\overline{\tau} \simeq0.8$ and
0.6 respectively).  Neither the
$\vy{\tau(v)}{S~VI}/\vy{\tau(v)}{Si~IV}$ nor the
$\vy{\tau(v)}{C~IV}/\vy{\tau(v)}{Si~IV}$ ratios for this feature are
approximately constant.  This behavior is indicative of large
variation in real $\tau(v)$ ratios since the S~VI and C~IV are
probably not strongly saturated. Ionization changes as a function of
velocity are the simplest explanation for non-proportional $\tau(v)$
once saturation is not the main culprit.

\section{EVIDENCE FOR NON BLACK SATURATION}

\subsection{Evidence for saturation and partial covering in PG~0946+301}

 \begin{figure}
\centerline{\psfig{file=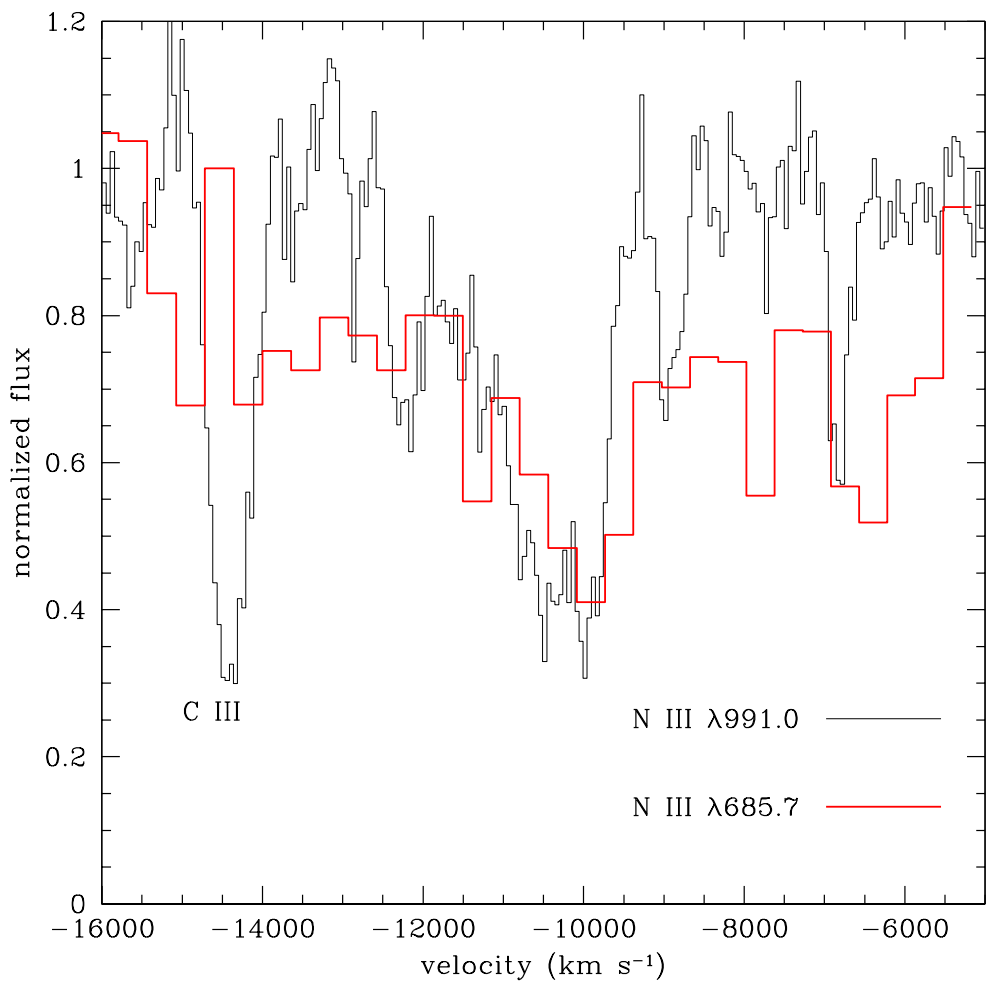,height=14.0cm,width=14cm}}

\caption{Residual intensity of the two N~III BALs presented on the
same velocity scale.  The multiplet at 685~\AA\ should have an optical
depth 2.3 times larger than the 991~\AA\ multiplet.  Although the data
for the 685~\AA\ BAL are of inferior quality, it is clear that the
apparent optical depth of both multiplets are equal within the errors.
This 1:1 ratio for the residual intensity in the two multiplets is a
direct evidence for non-black saturation in the BALs of PG 0946+301.
}\label{fig n3}
\end{figure}

\subsubsection{Optical depth ratio of the N~III BALs}

The best diagnostic for non-black saturation comes from comparing the
optical depth of two BALs from the same ion.  Transitions from the
same ion should have a $\tau$ ratio equal the ratio of their
$g_if_{ik}\lambda$ (where $g_i$ is the degeneracy of the level,
$f_{ik}$ is the oscilator strength of the transition and $\lambda$ is
the transition's wavelength). This $\tau$ ratio should not depend on
the ionizing flux, density, colisional processes, ionization
equilibrium or adundeces of the flow.  In the PG 0946+301 data  we
have two unambiguos occurences of two BALs from the same ion.  Two
O~III BALs (834~\AA\ and 703~\AA) are detected but the ratio of their
$g_if_{ik}\lambda$ is 1:1.1 and therefore they cannot be used as
saturation diagnostic since they are expected to have very similar
$\tau$. A good saturation diagonstic is found in the observed N~III
BALs (685~\AA\ and 991~\AA) which have
$(g_if_{ik}\lambda)_{685}/(g_if_{ik}\lambda)_{991}=2.3$. In figure 4
we show the two N~III BALs on the same velocity scale.  It is evident
that the residual intensities of component B in both BALs ($\sim
10,000$ \km) are the same within the quality of the data.  However,
based on the residual intensity of N~III~991~\AA\ (0.4) the
residual intensity of N~III~685~\AA\ should have been 0.12.

Before we can establish the ratio of apparent $\tau$ in the N~III BALs
as evidence for non-black saturation, we should examine
possible caveats.  If the absorption feature is significantly narrower
than the spectral resulotion, then the observed trough would be
shallower than expected.  Data of the N~III~685~\AA\ BAL were taken
with the G160L grating which has a FWHM spectral resolution of
$\sim1300$ \km.  Based on the wavelength and strength of the
N~III~685~\AA\ quadruplet transitions, we expect the FWHM of the
absorption feature associated with component B to be $\sim1200$
\km. Independantly, this width should be similar to that of the
same feature in N~III~991~\AA\, which also has a FWHM of $\sim1200$ \km.
Since the width of the expected feature and the resolution of the
grating are similar, the depth of the observed feature should not be
affected appreciably by instrumental-resolution.  Another possible
problem is the blending of the N~III~991~\AA\ BAL with the
C~III~977~\AA\ BAL.  If the C~III BAL contributes most of the optical
depth at the deepest part of the N~III~991~\AA\ BAL, it might explain
why the N~III~991~\AA\ trough is so deep compared with the
N~III~685~\AA\ trough.  Since the BALs are blended the only way to
test this hypothesis is to use results from the template fitting
technique (See \S~5).  From fitting the whole C~III trough with the
optical depth template extracted from the Si~IV BAL we find that at
the deepest part of the N~III~991~\AA\ BAL the $\tau$ contribution of
C~III is less then 0.2.  If we account for that contribution the $\tau$
ratio of N~III~685~\AA\ to N~III~991~\AA\ is 1.3 which is still
substentially different then the expected ratio of 2.3.

The availability of two BALs from the same ion allows to
solve for the real optical depth and the effective covering-factor
($C$) of the flow. $C$ is defined such that $(1-C)$ accounts for observed
 continuum-photons that are not
covered by the BAL flow and for photons that are scattered into the
observer's line of sight.  If scattering into the line of sight is
negligible, then $C$ is the continuum covering fraction of the
BAL flow.  The relationships between the residual intensity in each
line ($I_1$ and $I_2$), $C$ and the optical depth are given by:
\begin{eqnarray}
I_1-1+C&=&Ce^{-\tau} \\
I_2-1+C&=&Ce^{-R\tau}, 
\end{eqnarray}
where $R\equiv(g_if_{ik}\lambda)_{2}/(g_if_{ik}\lambda)_{1}$.  Given
$I_1$, $I_2$ and $R$, equations (1,2) can be solved numerically thus
enabling us to put limits on the degree of saturation seen in the BAL
flow.  The upper limit is achieved under the assumption that the part
of the C~III BAL that contributes to the absorption in component B of
N~III~991~\AA\ has the same $C$ as the latter.  Under such conditions,
since $I_1$ and $I_2$ of component B in the two N~III BALs are equal
within the error of measurement, the upper limit from equations (1,2)
is $\tau=\infty$.  When we add all the uncertainties due to photon
shot noise and continuum level for each line (see \S~5.2 for
discussion of continuum uncertainties), we find that
$\tau{\mbox{(B)}}_{\mbox{min}}=2.4$ based on 1~$\sigma$ deviation.
Since $\tau_{\mbox{apparent}}=1.0$, the minimal degree of saturation
(defined as $\tau_{\mbox{real}}/\tau_{\mbox{apparent}}$) in this case
is 2.4 and roughly 2.3 times larger for N~III~685\AA.  Assuming no
connection between the C~III absorption and component B of
N~III~991~\AA\ yields the minimal estimate for saturation since in
this case we subtract the C~III absorption contribution from the N~III
trough.  We obtain
$\tau{\mbox{(B)}}_{\mbox{min}}=2.9_{-1.4}^{+\infty}$.  In this case
the degree of saturation for N~III~991~\AA\ is $3.7_{-1.9}^{+\infty}$.
We conclude that the ratio of apparent $\tau$ in the N~III BALs is a
direct quantitative evidence for non-black saturation in the deepest
part of the BAL flow.

\subsubsection{The optical depth solution for the O~III 834~\AA\ BAL}

In figure 2 we show two solutions for $\tau (v)$ of O~III 834~\AA.  For
the upper solution we made the usual assumptions that the ratio of
optical depths for the different line-components equal the ratio of
their $g_if_{ik}\lambda$, and that the source is completely covered. It is
evident that this $\tau(v)$ is very different than those of all the
other lines and furthermore it has the largest negative values of
$\tau (v)$.  These two features indicate that our simple line
formation model of a homogeneous absorbing screen is wrong, and that
much of the structure seen is an artifact of the deconvolution
procedure being applied to a line profile that is not formed under the
assumed conditions.  In contrast, the lower solution assumes that the
absorption profile is caused by very optically thick material that
does not fully covers the continuum source. In this case all line
components cause an equal flux reduction, and hence can be described
as having equal effective $\tau$.  This $\tau$ should then be
interpreted as $- \ln (1-C)$.   The
similarity of the O~III $\tau$ profile of component B derived in this
way to those of Si~IV and C~III is a compelling evidence for non-black
saturation in component B.  On the other hand between $-8500 < v <
-4000$ \km\ the O~III solution with complete homogeneous covering
seems to be in a better agreement with the Si~IV solution than the
saturated O~III solution (especially components C and E).  This
suggests that the flow is not strongly saturated at that velocity
interval.

\subsubsection{Residual intensity in component B of Si IV and S VI}

Similar to the two N~III lines mentioned above, we can test for
saturation by comparing the residual intensity ($I_r$) of absorption
features seen in well separated doublets.  In PG~0946+301 the best
examples are the Si~IV and S~VI doublets.  Component B is well
separated in both doublets.  In Si~IV $I_r$(red)=$0.37\pm0.02$ and
$I_r$(blue)=$0.36\pm0.02$.  If component B was not saturated, then
based on $I_r$(red), $I_r$(blue) should have been 0.13. For S~VI we
obtain $I_r$(red)$=0.21\pm0.02$ and $I_r$(blue)=$0.17\pm0.02$ where we
would have expect $I_r$(blue)=0.04.  These two separate comparisons
are easily understood if component B is saturated.  However, this
evidence is not as strong as in the case of the N~III lines since the
width of the BAL trough is larger than the doublet separation for both
Si IV and S VI.  Therefore, other absorption might augment the optical
depth seen in the red component and indeed there is a valid
mathematical solution that can de-blend the doublet while assuming
non-saturation (the $\tau$ solutions presented for Si~IV and S~VI where
made under this assumption).  However, a saturated solution can be
computed just as well.  The argument here is statistical in nature. If
the absorption is non-saturated a special set of conditions must occur
in order for the red doublet line of component B to appear with the
same residual intensity as the blue doublet line. Furthermore, these
special circumstance are different for Si~IV and S~VI since their
doublet separation are not the same (1933~\km\ and 3558~\km,
respectively). This highly contrived scenario is avoided in a
straightforward way once non-black saturation is assumed.

\subsubsection{Similar $\tau$ for component B of the  low ionization lines}

Additional evidence for saturation and partial covering in component B
is the remarkable coincidence in which $\tau(B)_{max}$ in all four low
ionization lines is between 0.95--1.2 (see for example the troughs due
to C~III and N~III~991~\AA\ in Fig. 4).  To obtain an estimate for the
probability of such occurrence we make the simplest assumption that
each BAL can have an optical depth between 0.2--3 (a smaller $\tau$
will be difficult to detect and a larger one will be consistent with a
black profile) with equal probability.  The quality of the data allow
us to differentiate between at least 10 $\tau$ values separated by 1.3
multiplicative factors in the range 0.2--3.  Combining these two
assertions we find that the chance probability of having
$\tau(B)_{max}$ in all four low ionization lines is between 0.95--1.2
is less then $10^{-3}$.  This is therefore the probability that the
apparent $\tau(B)_{max}$ in these BALs is indeed the actual $\tau$.
Once again, these remarkably similar $\tau$ values are
naturally explained if component B is highly saturated but not black.

The larger apparent optical depth seen in the high
ionization lines is explained by a larger covering factor for this
gas, similar to the situation observed in Q0449--13 (Barlow
1997). Saturation probably also occurs in the broad component between
--9000 \km\ and --4500 \km\ where $\vy{\tau(v)}{O~VI}\simeq1.5$ and
roughly constant.  

 We conclude that non-black BAL saturation is prevalent in PG
0946+301.  The unequivocal evidence is seen in
component B by comparing the two N~III BALs (\S~4.1.1). Supportive
evidence comes from the optical depth solution of the O~III BAL
(\S~4.1.2), from the equality of residual intensity of the blue and
red doublet components in Si~IV and in S~VI (\S~4.1.2), and from the
remarkable coincidence of similar optical depth for component B in all
four low ionization lines (\S~4.1.4).

\subsection{Evidence for non-black Saturation in other objects}

Evidence for non-black saturation in the BALs has been accumulating in
the past several years.  Due to the crucial impact that non-black
saturation can have on BAL studies it is important to establish its
existence.  Therefore, besides our findings from the spectrum of PG
0946+301, we also describe evidence for non-black saturation from
other objects.

\begin{enumerate}

\item  Narrow intrinsic absorbers:

In several observed cases where the intrinsic absorption is
narrow enough to show a resolved doublet, we see non-black saturation
by comparing the optical depth ratios of the two resolved components.
Some of the examples are: BALQSO CSO 755 (Barlow \& Junkkarinen 1994),
UM 675 (Hamann et al. 1997) Q0449-13 (Barlow 1997), second trough of
BALQSO 0226--1024 (K92).  
BALs were defined as absorption troughs wider then 2000 \km\ (Weymann et
al. 1991) for the sole purpose of distinguishing them from intervening and
associated absorption systems.  Narrower intrinsic absorbers share all
the physical character of `classical BALs': Variability on $\sim$
year time scale, smoothness of the absorption trough, similar
ionization state.  For these reasons the narrow intrinsic absorbers
should be regarded as the narrow-velocity-width extension of the BAL
phenomenon.  Furthermore, there is no gap in the distribution of
velocity width in intrinsic absorbers.  Such systems can be found
across the whole range from a few hundred to a few ten-thousands \km\
width.

In `classical BALs' the width of the trough is larger than the
observed doublet separations and therefore establishing cases of
non-black saturation is not as simple.  However, since non-black
saturation is evident in narrow intrinsic absorbers, it seems prudent
to hold that wider intrinsic absorption (i.e. BALs) are also saturated
until proven otherwise.

\item BALQSO 0226--1024:

Using the measurements of K92 and Turnshek et al. (1996), Arav (1997)
demonstrated that in BALQSO 0226--1024 the optical depths in the C~IV,
N~V and O~VI BALs are identical within measurement errors. If we make
the simplest assumption that each BAL can have an optical depth
between 0.3--3 (a smaller $\tau$ will be difficult to detect and a
larger one will be consistent with a black profile) with equal
probability, then the chance occurrence of this coincidence is less
than 1\%.  Non-black saturation is a simple non-contrived explanation
for this coincidence.

\item Comparison with intervening absorption:

BALs are very rarely black.  In the sample of 72 BALQSOs taken by
Korista and Weymann (Korista et al. 1993) there are only two objects
(Q0041--4032 and Q0059--2735) that have zero flux in the C~IV BAL
across three or more resolution elements.  Only one object
(Q0059--2735) shows a similar black Si~IV BAL (out of 69 objects that
cover the appropriate wavelength interval), and no object shows a
black N~V or \Ly\ BALs (out of 58 objects that cover the appropriate
wavelength interval).  This is very different from intervening absorption
systems where high resolution observations show a considerable
fraction of absorption troughs to have zero intensity at their
bottoms. Furthermore, wider intervening absorption systems are much
more likely to show zero intensity in both \Ly\ forest lines and metal
lines.  With no physical justification, the odds that only three
out of 191 observed BALs will reach zero intensity are negligible
(less than $10^{-51}$, if we make the naive assumption of equal
probability for black and non-black troughs).  Therefore, unless  a
compelling reason for the absence of black BALs is found, the extreme
rarity of black BALs is a strong indication for non-black saturation
in most BALQSOs.

\item Spectropolarimetry observations:

 Spectropolarimetry of BALQSOs provides direct evidence for photons 
filling in the bottoms of the BALs. Cohen et al (1995) and Ogle (1997) show
that in most objects where they obtained spectropolarimetry data, the
polarization at the bottom of the troughs is much higher than the
continuum level polarization.  This finding demonstrates that a considerable
flux at the bottom of the troughs is from scattered photons.
Therefore, a known and verified mechanism is filling in the
bottoms of the troughs and can easily cause non-black saturation.

Ogle (1997) study of spectropolarimetry data in BALQSO 0226--1024
deserves a special mentioning.  Due to its width and depth, the broad
high velocity trough of this object (--14,000 to --20,000 \km) is
definitely a proper BAL absorption.  Ogle finds that the polarization
level in this trough is about 7\% compared to 2\% in the continuum.
Since only a fraction of the scattered light is polarized,
it can account for much, if not all, of the residual
intensity in this trough. The morphology of the trough in polarized
light supports this assertion.  Instead of being flat bottomed 
(as it appears in non-polarized light) it shows considerable
variations in optical depth..  This contrast in
morphologies suggests a high degree of saturation in the direct light
path, which is therefore insensitive to moderate optical depth
fluctuations, whereas the smaller optical depth seen in polarized light
reflects these fluctuations readily.

\item SBS1542+541:

While revising this paper, we became aware of an extensive analysis of
the very high ionization BALQSO SBS1542+541.  In their paper Telfer et al
(1998) show unambiguous cases of non-black saturation in the observed
BALs.  Their analysis is facilitated by the fact that the BALs of
SBS1542+541 are less than 3000 \km\ wide and therefore the doublet
components of most lines from the lithium iso-sequence are fully
resolved (Ne VIII, Mg X and Si~XII).            

\end{enumerate}

\section{RESULTS FROM TEMPLATE FITTINGS}


As was found in BALQSO 0226$-$1024  (K92),
the spectrum below 1000~\AA\/ is rich in resonance line troughs. Trough
overlap and confusion with the Ly$\alpha$ forest make measurements of
individual BAL troughs difficult. While the Ly$\alpha$ forest in
PG~0946$+$301 does not confuse the spectrum nearly as much as in the
higher redshift BALQSO 0226$-$1024, we had to rely on synthetic spectrum
fitting using optical depth templates to identify and measure many
of the BALs.

\subsection{The Optical Depth Templates}
We began our analysis by isolating two relatively strong troughs that
are unblended with other BALs or broad emission lines: Si~IV and C~IV.
These served as our optical depth templates for creating the synthetic
spectra.  Using the method described in Appendix A, we derived the
optical depth of the stronger doublet transition as a function of
outflow velocity for each of these two ions. Essentially, we assume
the residual intensity is equal to $e^{-\tau}$, with the caveats
described in K92. For a choice of effective continuum, these caveats
ensure that we measure a lower limit to the ionic column density. The
effective continuum is defined as that which contains all sources of
photons which are scattered by the BAL flow; this includes the central
continuum source, and in some instances emission lines. Differences in
the optical depth velocity profile that cannot be accounted for by a
velocity-independent scale factor indicate the presence of ionization
stratification (see \S~3.2 and \S~4.1) and/or differences in the line-of-sight
source coverage.

\subsection{Deriving the Effective Continuum}
A modified version of a mean quasar emission line spectrum (Weymann et
al.\ 1991) used by K92 formed the basis of the emission line portion of
the effective continuum.  This spectrum was first normalized by a fit
to the underlying continuum. Then using Gaussians, additional emission
lines of O~III $\lambda$834 and O~IV $\lambda$789 $+$ Ne~VIII
$\lambda$774 were added.  Slight additions to the broad emission lines
of C~III $\lambda$977, O~VI $\lambda$1034, N~V $\lambda$1240, Si~IV
$\lambda$1400, and C~IV $\lambda$1549 were made, but otherwise the mean
quasar emission line spectrum fit the strong lines of PG~0946$+$301
quite well. A small narrow line contribution to Ly$\alpha$
$\lambda$1216, uncovered by the BAL outflow was also added to the
spectrum. This narrow feature is seen in the spectrum near 1216~\AA\/
and is almost certainly uncovered, since a tremendous EW in this narrow
line would be required to match the feature if it were covered by the
outflow with the optical depth of the N~V $\lambda$1240 resonance
line. Its rest frame FWHM is approximately 4~\AA\/ (985 \km\/) and has
an integrated rest frame flux of $\approx 8.5 \times 10^{-15}$
ergs/s/cm$^{2}$.

A 4-piece power law continuum was allowed to vary in the fit. The
pieces were pinned together at 855~\AA\/, 1060~\AA\/, and 1150~\AA\/,
and the fit was allowed to determine the overall normalization plus 4
power law indices.  However, the fitting-determined continuum is not
the best approximation for the true continuum due to the following
reasons.  First, our fitting routine did not account for intervening
absorption seen in the ``high'' quality data ($\lambda>800$~\AA\ restframe).
As a result the fit consistently underestimated the true
continuum level.  We have corrected for that by forcing the continuum
up by 5--10\% compared to the fitting-determined one.  Second, in the
``low''-quality data ($\lambda<800$~\AA\ restframe) the fitting routine tends
to find unrealistically high continuum due to the small size  of
BAL-free continuum segments and the low quality of data in this regime. Given
enough free parameters and narrow BAL-free continuum-segments, the
fitting routine tends to get better $\chi^2$ fit by greatly
overestimating the continuum level and correcting for that by
deepening the many available BALs.  Our optimal continuum for
this region is based on more physical criteria: It matches
the two BAL-free continuum-segments (see \S~5.7), it has only one free
parameter which seems desirable in view of the low data-quality, and
most of the flux above it can be associated with expected broad
emission lines (O~III~$\lambda$703, O~V~$\lambda$630 and
He~I~$\lambda$584).  Due to the low data-quality and since modeling
these BELs was not important in the BAL analysis, we opted not to
include them as extra free parameters in the model.

Our optimal continuum longward of 800\AA \ is between the one
found by the fitting algorithm and the one created independently by
Junkkarinenet al. (1997).  These continua, compared to the data, appear
to be lower and upper limits to the true continuum and from these
we estimate a conservative error of 10\% for our optimal continuum.
Shortward of 800~\AA\ we estimate the
error in fixing the continuum level to be 10\% at 800\AA\ 
increasing to 20\% around 600\AA.  For comparison, results of both the
optimal continuum and the fitting-determined continuum are presented in
Table 1.    

\subsection{Fitting the Spectrum}
The normalized line
spectrum was multiplied by the assumed continuum to form the
effective continuum that was used by the fitting algorithm. 
Scaled versions of the appropriate optical depth template were then
placed in the rest frames of all transitions included in the fit.
Atomic data for these transitions --- wavelengths, oscillator
strengths ($f_{ik}$), and statistical weights ($g_{i}$)--- were found
in Verner, Barthel, \& Tytler (1994) and Verner, Verner, \& Ferland
(1996).  Initially, a comprehensive metal line list was compiled to
include all possible transitions, lying within the spectral range of
the observations, of parent ions whose ionization potentials lay above
that of C~II (24~eV). This cutoff was imposed due to the lack of a
C~II $\lambda$1335 BAL.  Fits of the spectrum were then constructed
(described below), and gauging the oscillator strengths and the
ionization potentials of ions whose transitions were too weak to be
constrained by the data, certain ions were iteratively rejected. These
included N~II and Fe~III --- the two lowest ionization ions
considered.  We included He~I $\lambda$584 to search for the presence
of this important ion. The strongest transitions of each ion present
in our final fit are listed in Table~1, along with a code that
identifies which optical depth template was used to fit the ion's BAL
trough.  The S~IV transition listed in Table~1 is not the strongest
for reasons that will become clear in \S~5.7.

We used the C~IV template for the stronger, higher ionization troughs
and Si~IV for the weaker lower ionization troughs. This makes physical
sense and produced the best fits for the stronger lines. H~I is an
exception, since although it has a small ionization potential, some
neutral hydrogen is always available. We modeled it with both
templates. All transitions for each ion were considered; all were
scaled in strength to the strongest transition by
$g_if_{ik}\lambda\/$.  The optical depth template scale factors, or
``multipliers'' for every ion were optimized using a downhill simplex
minimization routine (``Amoeba''; Press et al.\ 1992).  A $\chi\/^2$
was computed using the statistical error bars of the data and
comparison of the synthetic spectrum to the observed one. Prior to
optimization the observed spectrum was normalized by a fit to the
effective continuum, described above. The optimal optical depth
multipliers were those that minimized $\chi\/^2$.  Once a solution was
found, the process was restarted from the new starting point --- a
single iteration proved sufficient to converge to the minimum.  A full
description of this process may be found in K92. The inclusion of the
effective continuum in the optimization did not change the results in a
believably significant manner.

\begin{figure}[htb]
\hspace{-2.5cm}
\centerline{\psfig{file=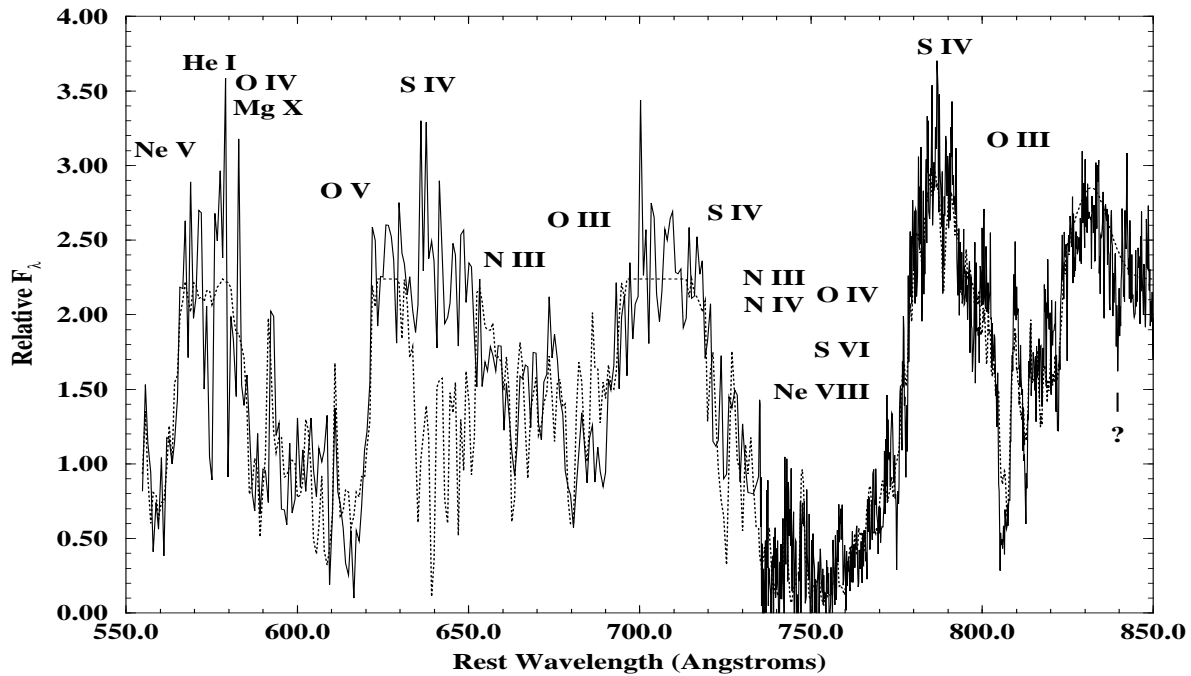,angle=-90,height=10.0cm,width=10cm}}

\caption{The solid line is the observed spectrum of
PG~0946$+$301 and the synthetic spectrum is dotted. The vertical axis
has the same units as Figure~1. Ions are labeled near their principle
troughs.  }
\label{fig 8}  
\end{figure}

 \begin{figure}[htb]
\hspace{-2.5cm}
\centerline{\psfig{file=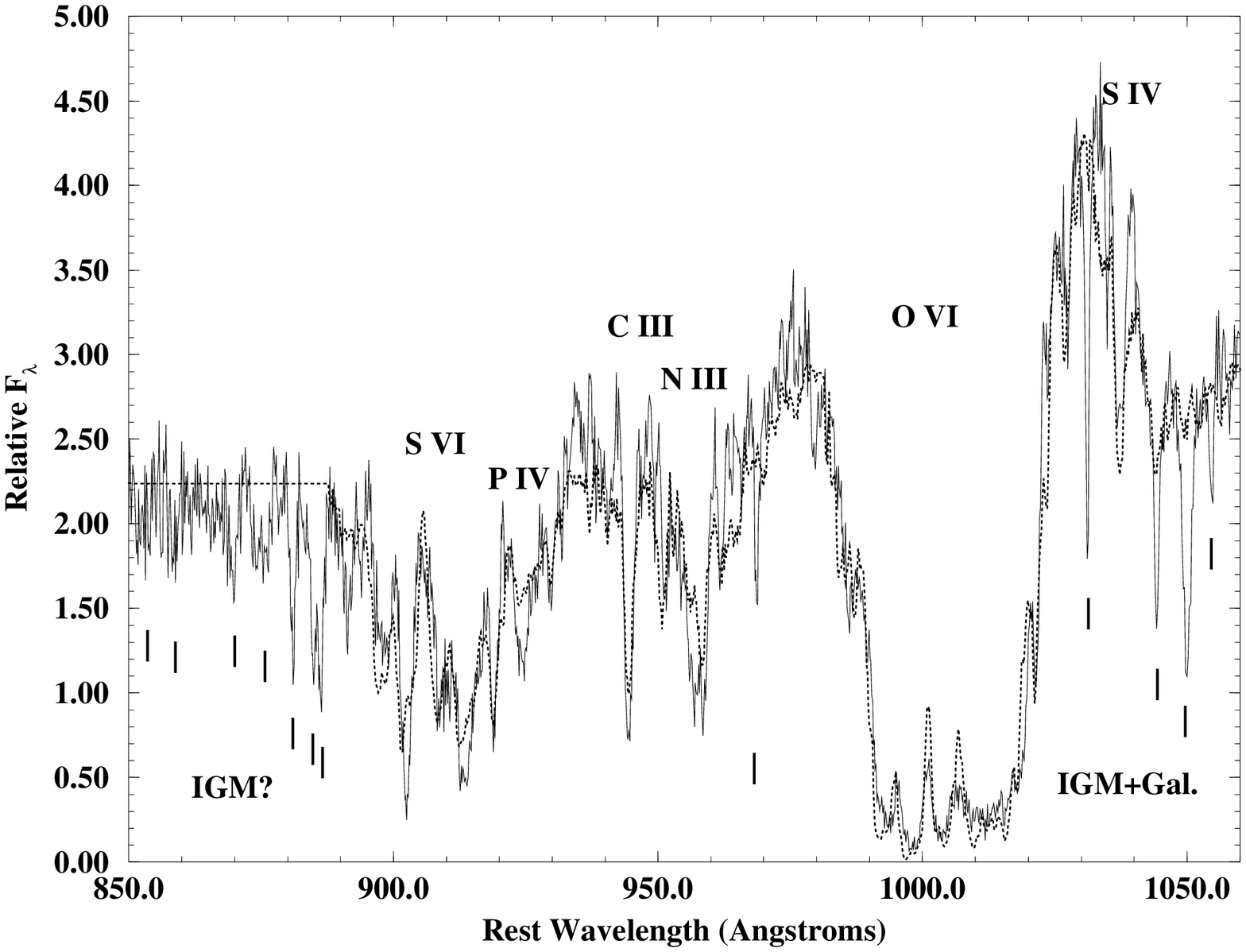,angle=-90,height=10.0cm,width=10cm}}

\caption{The solid line is the observed spectrum of
PG~0946$+$301 and the synthetic spectrum is dotted. The vertical axis
has the same units as Figure~1. Ions are labeled near their principle
troughs. }
\label{fig 9}  
\end{figure}

 \begin{figure}[htb]
\hspace{-2.5cm}
\centerline{\psfig{file=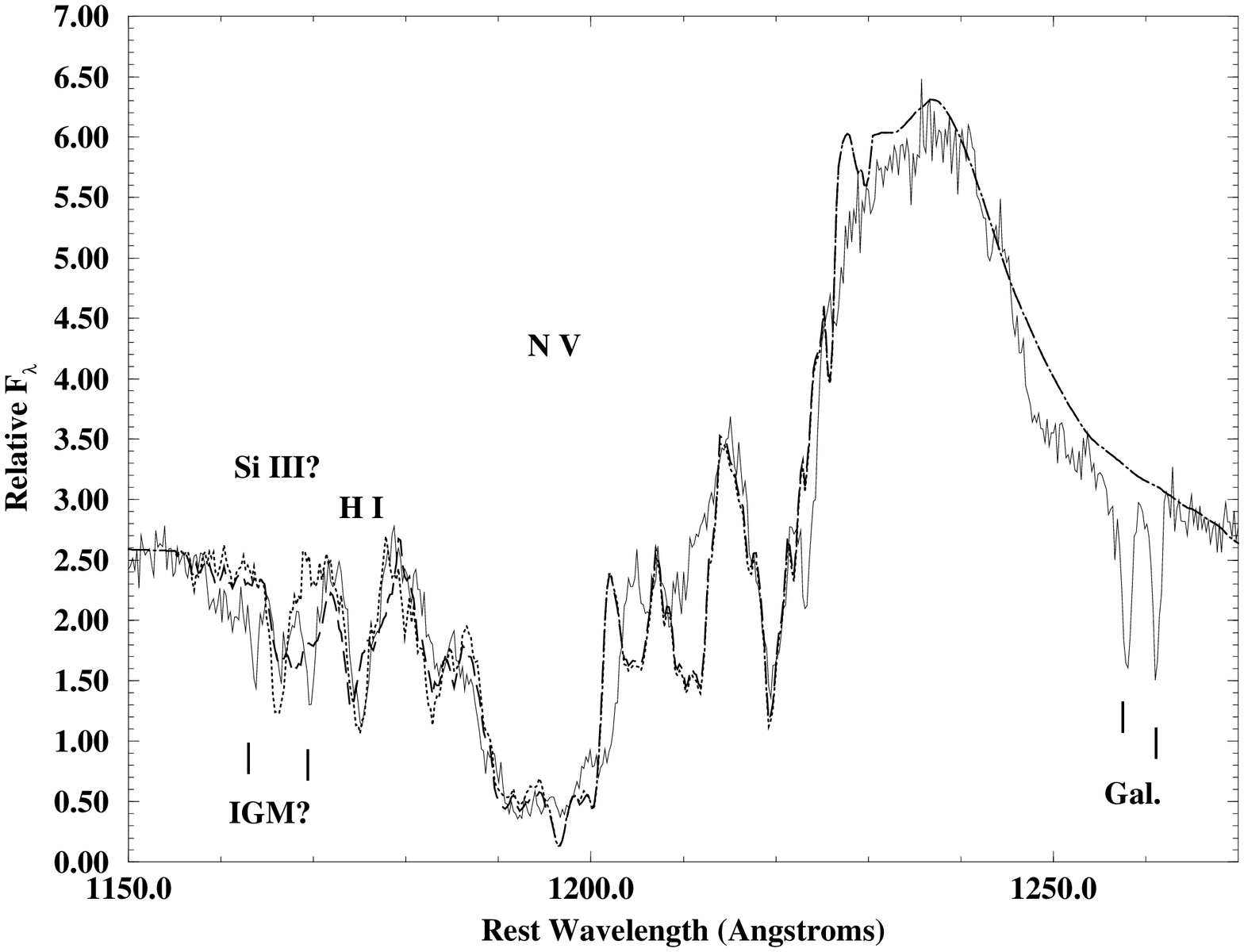,angle=-90,height=10.0cm,width=10cm}}

\caption{The solid line is the observed spectrum of
PG~0946$+$301 and the two synthetic spectra are dotted and dot-dashed. The
dotted one assumed the Si~IV optical depth template for the fit to the
BAL trough of Ly$\alpha$, the dot-dashed one assumed the C~IV template. The
vertical axis has the same units as Figure~1. Ions are labeled near
their principle troughs. }
\label{fig 5c}  
\end{figure}

In Figures~5--7 the synthetic spectrum of PG~0946$+$301 for the case
of a fixed effective continuum is plotted over the observed.  The
optical depth multipliers for the fixed continuum case and the derived
integrated ionic column densities for both types of effective continua
are given in Table~1. Also given in Table~1 are the velocity integrated
column densities computed directly from the 7 ionic optical depth
templates derived in $\S$~2.  To compute the column densities from
these templates we assumed:  
\begin{equation} \rm{N_{ion}} \geq \frac{
\rm{m_ec} }{ \pi\/\rm{e}^2 } ~ \frac{1}{\lambda_{\rm{o}}\rm{f_{ik}} }
\int {\tau\/(v) dv} = \frac{ 3.7679 \times 10^{14}~\rm{cm}^{-2} }{
\lambda_{\rm{o}}\rm{f_{ik}} } \int {\tau\/(v) dv}, 
\end{equation} with velocity
measured in \kms. In all cases but C~III and N~III, which are
contaminated with each other, the results derived from the fits match
well those computed directly. This result helps validate our method of
representing the BALs troughs of other ions by templates constructed
from the BAL troughs of C~IV and Si~IV. The C~IV and Si~IV templates
are not a perfect match at every velocity, however, as Figure~2 and
Figures~5--7 show. Except for C~III and N~III, the column densities
listed in column~6 of Table~1 should be the most accurate.

\pagebreak

\begin{table}[htb]
\begin{center}
\begin{tabular}{ccccccc}
\multicolumn{7}{c}{\sc Table 1: Apparent Ionic Column Densities} 
\\[0.2cm]
\hline
\hline
%
\multicolumn{1}{c}{Ion}
&\multicolumn{1}{c}{Transition (\AA\/)}
&\multicolumn{1}{c}{Template$^a$}
&\multicolumn{1}{c}{Multiplier}
&\multicolumn{1}{c}{$\rm{log N_{ion}^b}$ }
&\multicolumn{1}{c}{$\rm{log N_{ion}^c}$ }
&\multicolumn{1}{c}{$\rm{log N_{ion}^d}$ }\\
\multicolumn{1}{c}{(1)} & \multicolumn{1}{c}{(2)} & 
\multicolumn{1}{c}{(3)} & \multicolumn{1}{c}{(4)} &
\multicolumn{1}{c}{(5)} & \multicolumn{1}{c}{(6)} & 
\multicolumn{1}{c}{(7)} 
\\[0.05cm]
\hline
H I & 1215.670 & A & 0.36 & 15.32 & -- & 15.25 \\
H I & 1215.670 & B & 0.85 & 15.04 & -- & 14.99 \\
He I & 584.334 & B & 0.28 & 15.04:: & -- & 15.56:: \\ 
C III & 977.020 & B & 0.93 & 14.91 & 14.82$^e$ & 14.88 \\ 
C IV & 1548.195 & A & 1.00 & 15.99 & 15.99 & 15.99 \\ 
N III & 991.577 & B & 0.43 & 15.42 & 15.68$^f$ & 15.39 \\
N IV & 765.147 & A & 0.76 & 15.67 & -- & 15.77 \\
N V & 1238.820 & A & 1.28 & 16.28 & -- & 16.26 \\  
O III & 835.2891 & B & 0.97 & 15.93 & 15.99 & 15.88 \\ 
O IV & 790.199 & A & 0.35 & 16.12 & -- & 16.11 \\ 
O V & 629.732 & A & 1.17 & 16.02 & -- & 16.06 \\ 
O VI & 1031.926 & A & 2.08 & 16.64 & 16.61 & 16.63 \\ 
Ne V & 572.338 & A & 0.63 & 16.62 & -- & 16.77 \\ 
Ne VIII & 770.409 & A & 1.39 & 16.71 & -- & 16.69
\\[0.01cm]
\hline
\end{tabular}
\end{center}
\end{table}

\pagebreak

\begin{table}[h]
\begin{center}
\begin{tabular}{ccccccc}
\multicolumn{7}{c}{\sc Table 1 Continued: Apparent Ionic Column Densities} 
\\[0.2cm]
\hline
\hline
%
\multicolumn{1}{c}{Ion}
&\multicolumn{1}{c}{Transition (\AA\/)}
&\multicolumn{1}{c}{Template$^a$}
&\multicolumn{1}{c}{Multiplier$^b$}
&\multicolumn{1}{c}{$\rm{log N_{ion}^b}$ }
&\multicolumn{1}{c}{$\rm{log N_{ion}^c}$ }
&\multicolumn{1}{c}{$\rm{log N_{ion}^d}$ }\\
\multicolumn{1}{c}{(1)} & \multicolumn{1}{c}{(2)} & 
\multicolumn{1}{c}{(3)} & \multicolumn{1}{c}{(4)} &
\multicolumn{1}{c}{(5)} & \multicolumn{1}{c}{(6)} & 
\multicolumn{1}{c}{(7)} 
\\[0.05cm]
\hline 
Na IX & 681.719 & A & 0.14 & 15.80:: & -- & 15.90:: \\ 
Mg X & 609.7930 & A & 0.57 & 16.51 & -- & 16.60 \\ 
Si III & 1206.500 & B & 0.71 & 14.36:: & -- & 14.17:: \\ 
Si IV & 1393.755 & B & 1.00 & 14.95 & 14.95 & 14.95 \\
P IV & 950.657 & B & -- & -- & -- & --\\ 
P V & 1117.977 & B & -- & -- & -- & --\\ 
S IV & 750.222 & B & 1.85 & 15.41: & -- & 15.32:: \\ 
S V & 786.464 & A & 0.53 & 15.14 & -- & 15.12 \\ 
S VI & 933.378 & A & 0.63 & 15.64 & 15.63 & 15.62
\\[0.01cm]
\hline
\end{tabular}
\end{center}
$^a$A is C IV template; B is Si IV template.\\
$^b$Optical depth template multiplier and column density derived from
template fitting using a fixed effective continuum. (::) identification
and column density both highly uncertain; (:) identification secure,
but column density highly uncertain\\
$^c$Column density derived from direct optical depth solutions
using a fixed effective continuum.\\
$^d$Column density derived from template fitting 
using a floating effective continuum.\\
$^e$Velocity interval of integration 7000--14000 \kms\ to minimize
contamination by N~III.\\
$^f$Velocity interval of integration 7500--13000 \kms, but still
contaminated by C~III.
\end{table}

\pagebreak

\subsection{The Highest Ionization Lines: Ne~VIII, Na~IX, and Mg~X}

We report the presence of very high ionization BAL troughs from two
members of the Li isoelectronic series: Ne~VIII and Mg~X. K92 reported
the possible presence of Ne~VIII in 0226$-$1024 and Telfer et
al.\ (1998) have identified the BALs of Ne~VIII,  Mg~X and Si~XII in the
extraordinary SBS1542$+$541. The ionization potentials of these ions
are over 200~eV, and significant column densities would point to the
presence of highly ionized gas as an important constituent of the BAL
outflow.  The presence of such highly ionized gas is required by
some models for the formation of BALs (Murray et al. 1995, Murray \&
Chiang 1995), and measured column densities for these species are also
very important for models connecting the BAL absorption with the
highly ionized soft X-ray absorbers seen in some AGN 
(e.g. Mathur et al. 1994, Mathur, Wilkes \& Aldcroft 1997). 

In Figure~8 we show the results of removing Mg~X $\lambda$
$\lambda$610,625 from the fitting process. The steps above were repeated with
these transitions absent, and in this case we allowed the fitting
process to vary the underlying power law continua, giving it more
freedom. In the end, however, this had very little effect upon the
results below. The stronger transition of the doublet overlaps with
O~IV $\lambda$609. In Figure~8 one can see a large discrepancy
between the fit and the data in just the region where the largest
optical depths occur for the Mg~X doublet. The O~IV $\lambda$609
transition is not allowed to get stronger to deepen the synthetic
trough near 587~\AA\/ because its strength is constrained by that of
O~IV $\lambda$789. Even if it could, the region near 600~\AA\/ would
still be too high in the synthetic spectrum; the depression here is
due to mainly Mg~X $\lambda$625.  The fit shown in Figure~5, which
includes Mg~X, is a much better one. 

 \begin{figure}[htb]
\vspace{1cm}
\hspace{-2.5cm}
\centerline{\psfig{file=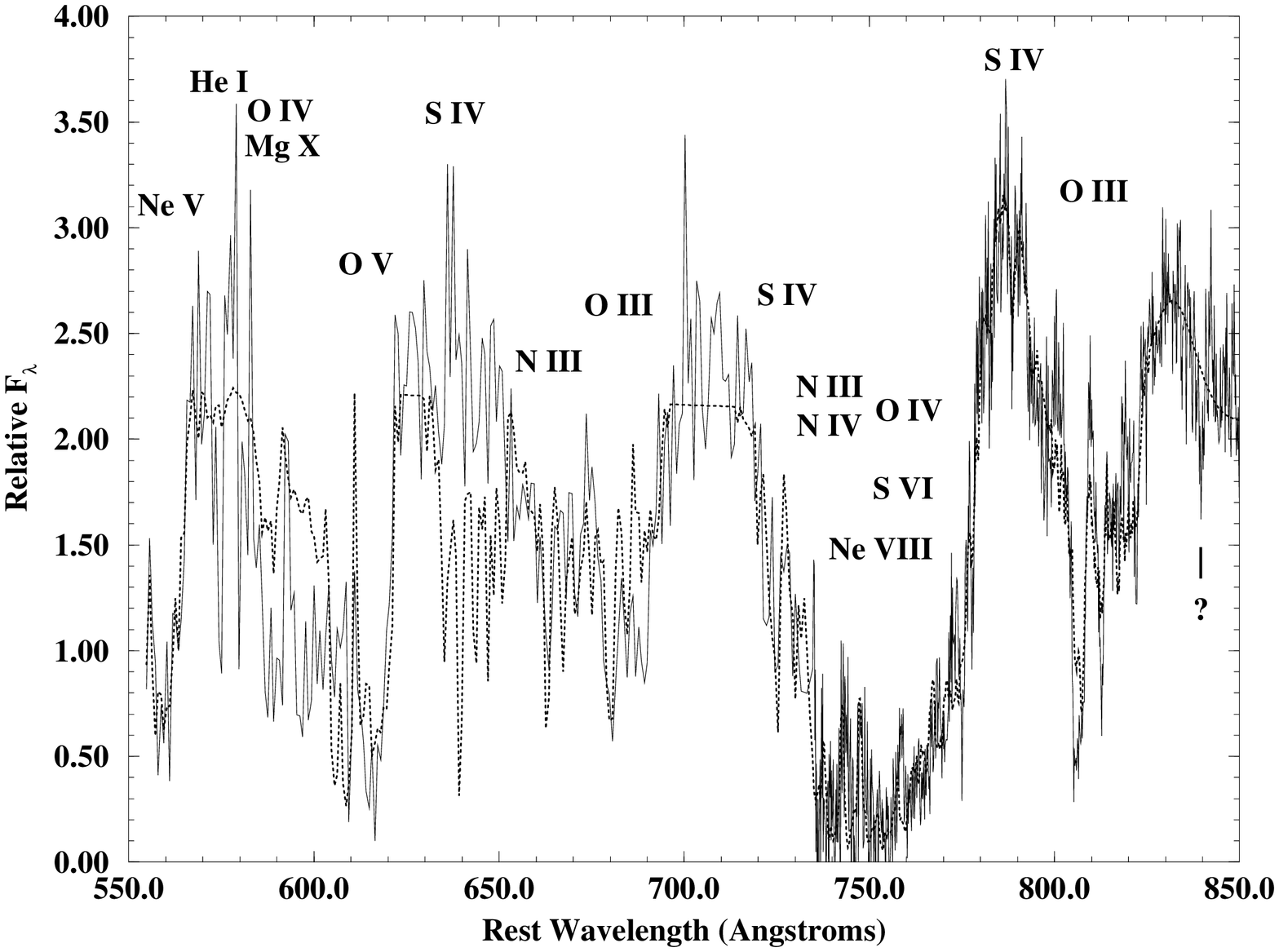,angle=-90,height=10.0cm,width=10cm}}

\caption{The same as Figure~8 but with the Mg~X doublet
removed from the fit. }
\label{fig 6a}  
\end{figure}

In Figure~9 we demonstrate the same effect of the absence of Ne~VIII
$\lambda \lambda$770,780 from the spectrum. Upon comparison with
Figure~5 it is seen that the derived synthetic spectrum allows too
much light too leak through the deep 730--780~\AA\/ feature, especially
near 742~\AA\/ and 763~\AA\/. The confirmed presence of Ne~V (the deep
trough at the short wavelength end of the spectrum) and the apparent
presence of Mg~X also seem to suggest a significant column density in
Ne~VIII.

 \begin{figure}[htb]
\vspace{1cm}
\hspace{-2.5cm}
\centerline{\psfig{file=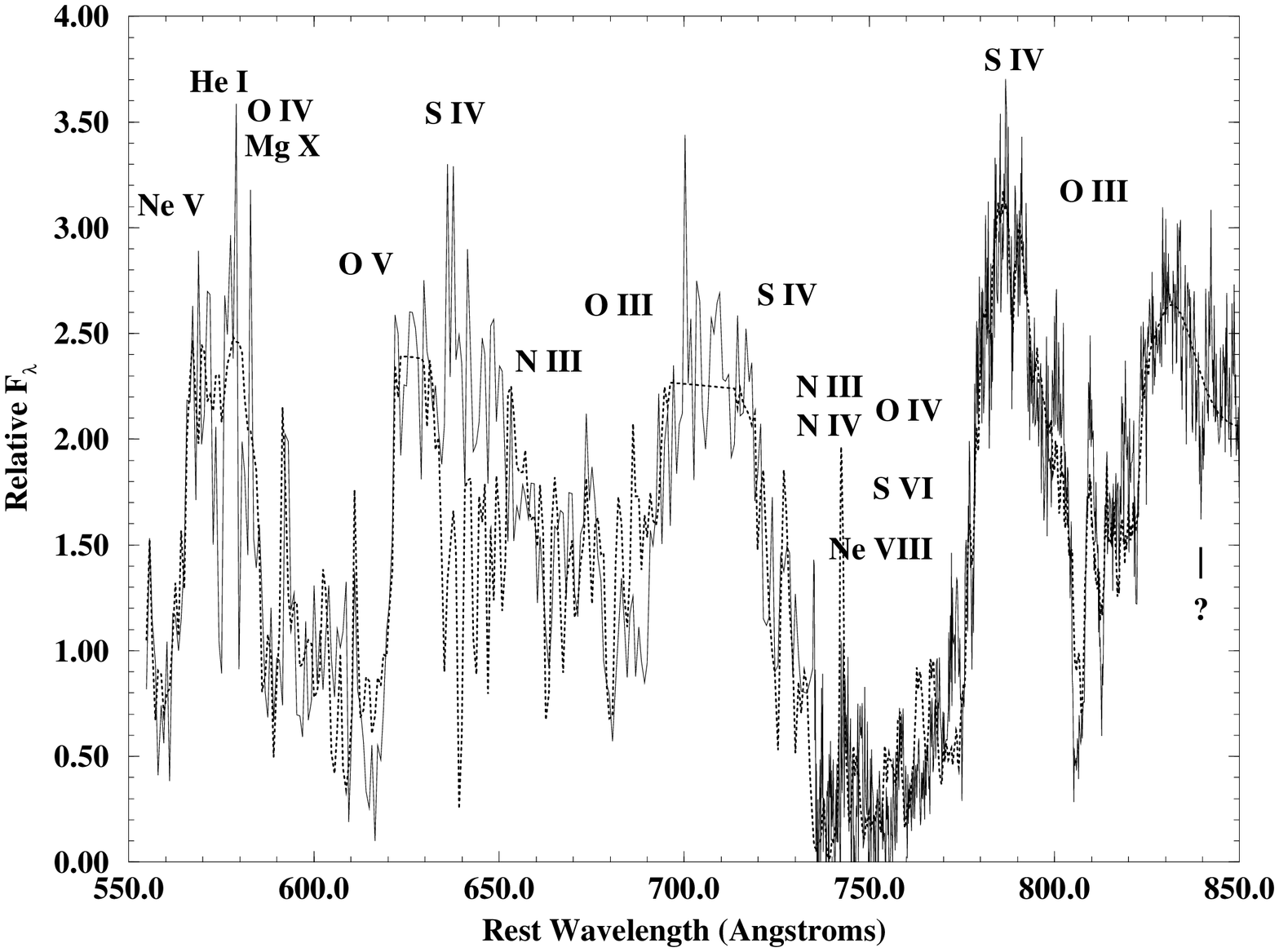,angle=-90,height=10.0cm,width=10cm}}

\caption{The same as Figure~8 but with the Ne~VIII doublet
removed from the fit. }
\label{fig 6b}  
\end{figure}

The transitions of Na~IX $\lambda \lambda$682,694 also occur in the
G160L spectrum. Their presence or not in the line list had too little
impact on the fit to determine the reality of this ion's presence. This
is because whatever might be there is apparently weak --- as found by
the fit, the multiplier for the stronger transition indicates an
optical depth of only 14\% that of C~IV $\lambda$1548 (Table~1). Higher
quality data will be necessary to determine the presence of the ion of
this normally under-abundant element.

\subsection{The Phosphorus Abundance}

The P~IV trough overlaps, unfortunately, with that of the weaker
transition of S~VI ($\lambda$944); it is not likely that we will ever
be able to measure its possible contribution in this object.  However,
we confirm the identification of P~V $\lambda$1118,1128 reported by
Junkkarinen et al. (1997). We do not report the analyses of this spectral
region here; however, in Figure~1 one may view the depression in the
continuum near 1100~\AA\/ due to P~V. The fit to this spectral region
and the derived column density of P~V will be presented in Junkkarinen
et al.\ (1998).

\subsection{He~I and Si~III}

Given that helium may be a significant opacity source in the flow, it
would be important to set meaningful limits on its neutral column
density. A measured column density in a neighboring ion to Si~IV would
help constrain silicon's ionization balance.  If present at all, the
lines of He~I $\lambda$584 and Si~III $\lambda$1207 are apparently weak
and the present data do not allow for their solid identifications. The
troughs centered near 575~\AA\/ and near 1167~\AA\/ in Figures~5 and
7, respectively, are the features that the fit is identifying with
these two ions. The fits to both sets of features are rather poor.
Higher S/N, spectral resolution data to replace the G160L spectrum will
be necessary in confirming the presence of He~I; coincidental
intervening absorption may be confusing the fit.  The identification of
Si~III may also be confused with intervening absorbers as well as high
velocity Ly$\alpha$ BAL.
In using the C~IV optical depth template to fit the Ly$\alpha$ BAL, the
predicted Si~III column density fell to 46\% of its already small value
shown in Table~1. This is because the C~IV template has larger optical
depths at high velocities. He~I and Si~III are the lowest ionization
species remaining in the line list (excepting hydrogen) and probably
 are not important constituents of the BAL flow.

\subsection{The S~IV Conundrum}

Several sets of S~IV transitions occur throughout the spectrum, as
labeled in Figures~5 and 6. Significant absorption features that
appear on top of the O~VI $\lambda$1034 ($\lambda \lambda$1063,1073)
and the O~IV $\lambda$789 $+$ Ne~VIII $\lambda$774 emission lines
($\lambda \lambda$810,816) as well as the features on the short
wavelength edge of the broad depression near 730--780~\AA\/ 
(745~\AA, 748~\AA, 750~\AA, 754~\AA) together provide substantial evidence for
the presence of this ion. The last of these is particularly
interesting, since in the absence of S~IV the synthetic spectrum
recovers from the deep broad depression much more rapidly to shorter
wavelengths than is observed.  This problem was noted in the analyses
of BALQSO 0226$-$1024 (K92), but this set of transitions was
overlooked.

With the inclusion of S~IV all three of the aforementioned features are
fit reasonably well, but {\em only if} the oscillator strengths for the
weak doublet $\lambda \lambda$1063,1073  are increased over
their tabulated values by a factor of 2.7 (see Figs. 5 and 6). While
these lines are weak (tabulated $f_{ik} = 0.0434, 0.0487$), there is no
reason to suspect the accuracy of their oscillator strengths.  Morton's
compilation (1991) found similar values to those used here, derived
from the Opacity Project data base (Seaton et al.\ 1992; Verner et
al.\ 1996), and the oscillator strengths of the stronger transitions
should be more accurate. And yet, if the troughs of the weak
transitions ($\lambda \lambda$1063,1073) are fit, all of the rest are
predicted to be too strong by roughly the factor of 2.7.

However, even allowing for this ``correction'' the troughs of the
strongest S~IV transitions near 657~\AA\/ and 661~\AA\/ ($f_{ik}
\approx 1$) are not at all fitted. In fact, the effective continuum
level around 640~\AA\/ would have to lie 50\% -- 100\% higher than our
adopted level, and this excess would have to be  completely
uncovered by the outflow (see Fig~5). 
No strong emission lines are known to lie in
this region, and why would such emission be completely uncovered in any
case? Given the two spectral segments of BAL-free continuum at
625~\AA\ and 700~\AA\/ and the expected paucity of Ly$\alpha$ forest
lines at these redshifts, the emission line free continuum level near
640~\AA\/ could lie no higher than 20\% above our adopted level.
However, emission lines of O~V 630 and O~III 703 may account for the
flux in excess of our adopted continuum in these two BAL-free windows,
and so our adopted continuum level is probably accurate. Finally, only
2 of the 5 major troughs expected from the S~IV actually align with
corresponding features in the data. The possible absence of significant
S~IV is puzzling given the presence of other sulphur ions as well as
the fact that the creation and destruction ionization potentials of
S~IV are similar to those of Si~IV, C~III, and N~III.  The resolution
of this conundrum will have to await higher quality data.

\subsection{The N~V -- Ly$\alpha$ troughs}

The modeling of the N~V trough (Figure~7) is complicated by the fact
that an unknown broad-emission line profile of Ly$\alpha$ has been
scattered by the N~V ions in the BAL flow. A remnant of this strongest
of quasar broad emission lines is seen near 1216~\AA\/ in Figure~7.
This narrow feature is apparently uncovered by the BAL flow, as
discussed above. Also, it is not clear which if either of the optical
depth templates is best suited to modeling the Ly$\alpha$ BAL. We
decided to use both; the resulting fits appear as dotted and dashed
lines in Figure~7. Many features in this spectral region are matched,
others are not or are shifted in wavelength. In some places the Si~IV
template fits better, in others the C~IV template. Luckily, the bulk of
the N~V trough occurs $\sim$~5200~\kms\ shortward of the Ly$\alpha$
emission line peak, although a small contribution from the Ly$\alpha$
BAL overlaps here. Thus the fit to the N~V BAL does not depend heavily
on our choice of the Ly$\alpha$ broad emission line profile.  Also, the
choice of template for the Ly$\alpha$ fit had very little impact (3\%)
on the predicted N~V integrated column density, although it did affect
the inferred H~I column density by roughly a factor of 2. Both sets of
H~I column densities are listed in Table~1. The size of this difference
was the exception, not the rule, in the dependence of the determination
of the ions' column densities to the optical depth template used in the
fitting.

\section{Kinematic Model of the Flow}

A fundamental issue in the study of BAL flows is determining their
geometry and origin (Weymann 1997; de Kool 1997).  No compelling
evidence in favor of a specific picture exists and the uncertainty in
these issues is hampering our attempts to obtain a complete physical
model for the flows.  One of the main models is that of a disk wind
(de Kool and Begelman 1995; Murray et al. 1995) .  A disk wind gives a
natural origin for the ejected material and can explain the prevalence
of detached and multiple troughs seen in the BALs, which radial
flow-models are hard pressed to explain (Arav 1996, \S~2.3; Arav \& Li
1994).

A cylindrically-symmetric disk-wind is ejected upward from the surface
of the disk and is radiatively-accelerated radially once exposed to
the strong continuum source, thus producing curved stream lines.  Flow
sheets develop in the wind in one of two ways. Either the ejection of
the wind is restricted to certain annuli of the disk, or else, stream
lines from a whole disk wind do not stay parallel to each other but
tend to clump in several regions and effectively produce separate flow
sheets.  Numerical solutions of disk winds tend to show the latter
behavior (Stone 1998, private communication). Cylindrical symmetry of
the wind is necessary to explain the stability of observed features
against Keplerian motion and is natural consequence of a disk scenario.

We now describe a kinematic model specific to PG 0946+301, which is
based on the general picture described above.  We note however, that
although our model is consistent with the data other models cannot be
ruled out.  As seen in the highest ionization line (O~VI), the flow
has three main components: A, B and the broader component between
--9000 \km\ and --4500 \km\ (hereafter, the broad component).  We
argue that these components represent three separate outflows.
Supporting this assertion is the uniqueness of each component:
Component A has a higher ionization state than the other two (absence
of low ionization lines).  Component B is very similar in all the low
ionization lines, is heavily saturated and in O~VI it covers at least
93\% of the continuum source. This is significantly larger coverage
than is seen in the broad component, where the constant
$\vy{\tau(v)}{O~VI}$ suggests a saturated flow with effective covering
factor of 80\%. Furthermore, the broad component shows low ionization
BAL flow in subcomponents C, D and E and therefore is qualitatively
different than component A.

 \begin{figure}
\centerline{\psfig{file=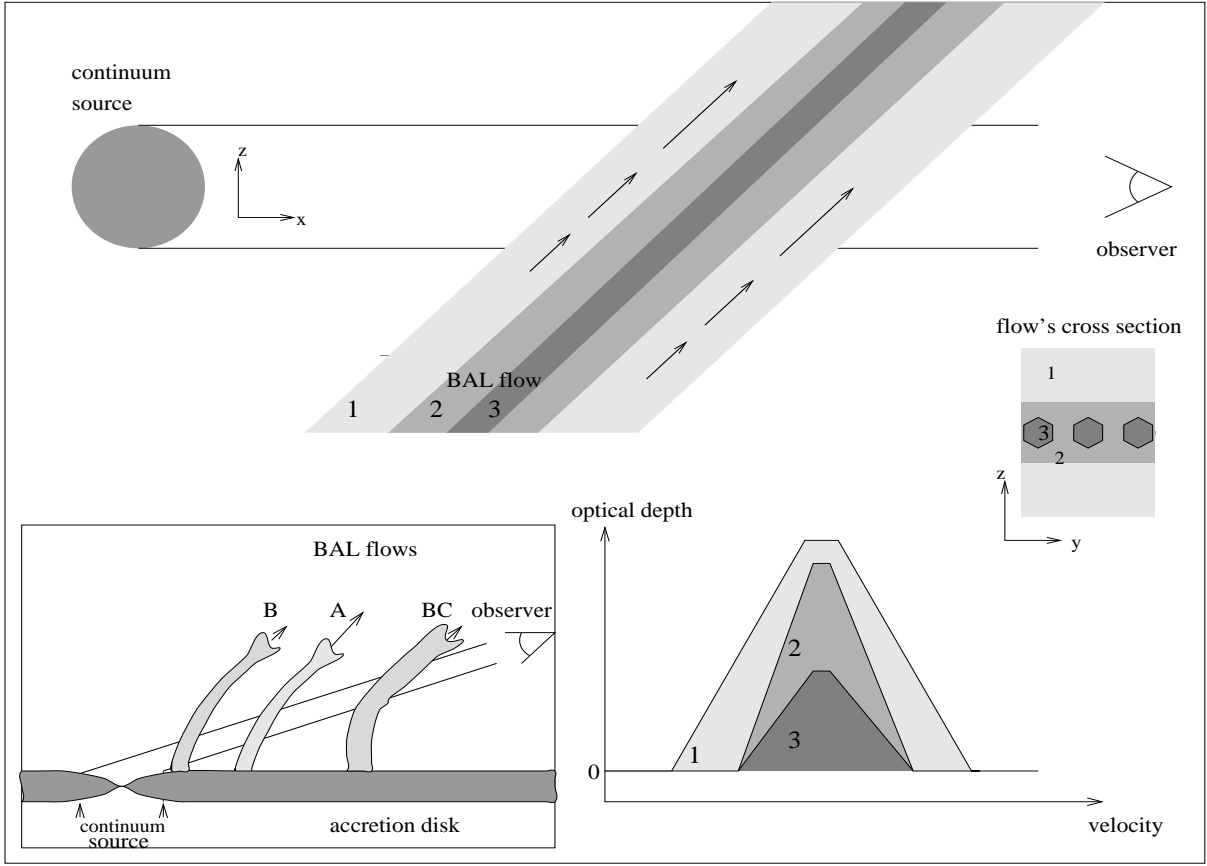,angle=-90,height=11.50cm,width=16cm}}

\caption{Illustration of the kinematic model. Top sketch shows an
accelerating BAL flow crossing the line of sight between the observer
and the continuum source, this geometry  explains the
appearance of stable (velocity-wise) detached troughs seen in most
BALQSOs. In PG 0946+301 we postulate the existence of three substrates
in the flow: 1. High ionization material with a small total column
density per unit velocity [$N(v)$]. 2. High ionization material with
 large $N(v)$.  3. Low ionization material embedded in
substrate 2 but with a smaller covering factor.  
On the middle right
of the figure we show a possible cross section for the flow
which assumes that the low ionization material appears in separated
flow tubes.  This gives the low ionization material smaller covering
factor than that of the high ionization material.  
In the plot at the bottom right of
the figure we show the resulting apparent optical depths from this model.
Substrate 1  represents the optical depth seen in the O~VI and
C~IV lines, substrate 2 represents the optical depth seen in the S~VI
and substrate 3 is for the low ionization lines O~III, Si~IV, N~III
and C~III. The cartoon at the bottom left shows a possible
geometry for the entire BAL flow seen in PG 0946+301 (BC: broad component).}
\label{kinematic}  
\end{figure}

Component B is seen in all seven lines and has a well defined profile,
therefore we concentrate on constructing a detailed flow model for it.
The model is based on the generic BAL flow-model described in Arav (1996).
What is seen in the HST data is a wide (velocity-wise; with similar
use of the term `wide' hereafter) high-ionization flow that reaches
$\tau_{max}\simeq2-2.5$ at --10100 \km.  In the midst of this flow
there is a low ionization component which is narrower and although it
is probably optically thick, it reaches only $\tau_{max}\simeq1.1$
(also at --10100 \km).  However, to complicate this picture, the high
ionization S~VI seems to follow the morphology of the low ionization
lines but with $\tau_{max}\simeq2$.  The simplest model that is
consistent with this phenomenology is of a wide high-ionization flow
with a positive gradient of total column density per unit velocity
[$N(v)$] toward its center (--10100 \km).  We propose that once $N(v)$
is high enough, sub-flows of lower ionization material are forming.
The low ionization flows have a lower covering factor than the
high-ionization flows, and one can picture them as flow tubes directed
along the high ionization flow sheets (see Fig. 4).  We invoke the
gradient of $N(v)$ in the high-ionization flow in order to explain the
absence of a wide S~VI component. Since the solar abundance of sulphur
is 20 and 50 times lower than those of carbon and oxygen, a larger
high-ionization $N(v)$ is needed to detect the S~VI line.  A larger
high-ionization $N(v)$ at the midst of the flow also makes the
appearance of low ionization material more plausible.  A smaller
covering factor for the low ionization flow is necessary once we
accept that the low ionization flow is saturated but not as deep as
the O~VI or S~VI flow.

The C~IV ground-based observation is of much lower resolution.  It
shows a well defined component B at the expected velocity position
with a FWHM of 800 \km\ and $\tau_{max}\simeq2.3$ (see Fig 2). An 800
\km\ FWHM value is  between the 1100 \km\ FWHM of O~VI and the
500 \km\ FWHM of the low ionization lines. In the context of the
kinematic model, we would expect such a behavior since carbon is less
abundant than oxygen (by a factor of 2.4) and since C~IV ionization
potential is 64 eV compared to 138 eV of O~VI.

Another feature that is widest in 
O~VI, less wide in C~IV and narrower still in the other lines,
 is the optical depth rise around --4000 \km. The
O~VI trough starts rising at --3500 \km\ and at --3700 \km\ it reaches
$\tau=1$ The C~IV trough starts rising only around --4000 \km\ and
reaches $\tau=0.5$ at --4200 \km. All the other BALs (including S VI)
have $\tau=0$ between --3800 \km\ and --4000 \km. Only beyond --4200
\km\ the other lines start having appreciable optical depths.  This is
consistent with the gradient of $N(v)$ we invoked above and again
shows a somewhat wider velocity covering for O~VI relative to C~IV.

To explain the smaller covering factor observed in the low ionization
flow, the model of component B  invokes the existence of low
ionization flow tubes within the high ionization flow sheets, A
prediction of such a model is that stronger variability should be
observed in the low ionization lines as the flow tubes move across the
line of sight in a keplerian motion.  However, this is not the case if
the number of flow tubes is large enough to smooth the variability
caused by the few flow tubes which are entering or leaving the line of
sight.

An alternative explanation for the different covering factors seen in
component B can be related to size of continuum emitting region.  If
the continuum emission arise from an accretion disk, the higher energy
photons should come from a smaller hotter part of the disk. Therefore,
the shorter wavelength lines in the flow should see a smaller emitting
region and thus will have larger covering factors.  We reject this
alternative since we would expect that the O~III line, which sees the
hardest continuum of all seven BALs, to have the highest covering
factor and thus the largest optical depth.  As the data show, this is
not the case.

Models for the other flow components are less constrained by the
data.  However, the main features of  component B model are consistent
with the data for the other two.  In the broad component we again have
a wide high ionization flow (seen in O~VI, C~IV and S~VI) in which low
ionization sub-flows are embedded (subcomponents C, D and E). Due to
the different morphologies, it is plausible that the C~IV and S~VI
absorptions are not saturated, whereas the O VI is.  This factor will
be significant for ionization and abundances studies of the flow. 
Higher quality data are needed to verify this assumption and for
studying the low ionization subcomponents in detail.  It is of course
possible that some of the subcomponents C, D, and E are in fact
totally separate outflows.  Component E has the highest chance of
being a separate flow since it is seen in all seven lines and is the
lowest velocity feature in the whole flow.  In component A only the
high ionization flow is present and the non (or marginal) detection of a
low ionization component might be correlated with the lower
$\tau_{max}(S~VI)$ seen in this component compared with
$\tau_{max}(S~VI)$ seen in Component B.

\section{DISCUSSION}

\subsection{The validity of the integrated column density approach 
in IEA studies}

Almost all attempts to study ionization equilibrium and abundances
(IEA) of BAL outflows used {\it apparent} integrated ionic column
densities ($N_{ion}$) measured from the BALs as the basis of the
analysis (Korista et al. 1996; Turnshek et al. 1996; Hamann 1996).  
It was always stated that these $N_{ion}$
are only lower limits due to the assumption made in their extractions
(see discussion in K92), but the severity of the problem was not realized
 and the IEA studies proceeded on the assumption that the
apparent $N_{ion}$ are a good approximation to the real
$N_{ion}$. The work presented here demonstrates the inadequacy of this
approach and thus strengthens similar conclusions reached by Arav (1997).  
As shown in \S~3.2 there are strong variations in the ratio of
apparent $\tau(v)$ in different BALs.  This feature suggests 
that comparing integrated $N_{ion}$ of different lines is
misleading and that it is necessary to work with $N(v)_{ion}$ to
obtain physically meaningful results.  Even more important is the 
evidence for BAL saturation over parts of their velocity extent
(\S~4.1).   Although the apparent $\tau_{max}$ in component B only vary
by a factor of two between the different BALs, 
the true optical depth can vary by a much larger factor (even a factor of
a hundred cannot be excluded), which is different for each ion.
 Therefore,  saturation and velocity-dependent ionization 
changes,  renders an IEA analysis based on 
apparent $N_{ion}$ futile and misleading.

 How can we use the available data
to obtain better constraints on the IEA in PG~0946+301?  From the
section above it is clear that before any ionization codes can be used
we must obtain a reliable determination of real $N(v)_{ion}$.  
Parts of the flow that are saturated can only tell about the geometry
of the flow.  On the other hand the broad
component (between --9000 and --4500 \km) is probably the least
saturated part in most BALs (see \S~3.2).  We can use the fact that O~VI
is saturated across the broad component in order to derive more
accurate $\tau(v)$ for the other lines.  While doing so we must
address the possibility that the covering factors of the ions may
differ.  We can make the assumption that the relative covering factors in the
broad component are similar to those seen in component B.  This is of
course by no means certain, only plausible. 
 Extrapolating from component B we can expect the covering
factors of C~IV to be most similar to O~VI followed by that of
S~VI.  Component B of S~VI is only half as wide as the
 O~VI one, but has a similar $\tau_{max}$.  In the broad component
it seems that the O~VI profile is broadest followed by C~IV and than
S~VI. However, this affects only about 1000 \km\ interval at the low velocity
edge of the flow (between --3500 and --4500 \km) and much less than
that at the high velocity edge. Therefore, it is reasonable to assume
that between --9000 and --4500 \km\ the covering factor of O~VI, C~IV
and S~VI are similar.  In the low ionization lines the broad component
does not appear to be saturated.  Therefore, assuming the same covering
factor as that of the high ionization lines should not lead to large
underestimates of real optical depth.

We note that the available data for the O~III~$\lambda$702,
O~IV~$\lambda$609, O~V~$\lambda$630 and N~III~$\lambda$685 BALs is of
too low a quality for IEA studies.  It is the combined analysis of
high quality data for these lines together with similar data for the
longer wavelength CNO BALs that will enable us to substantially
improve our understanding of the IEA in this object, and by extension
in high ionization BALQSOs in general.

\subsection{On the question of meta stable lines}

Pettini and Boksenberg (1986) identified the O~V$^*$ $\lambda$760.4
meta stable line in the IUE spectrum of PG~0946+301. Detection of meta-stable
lines is of high interest since these lines can only come from a very
high density gas ($n_e\gtorder 10^{11}$; see K92 for discussion).
Therefore, we checked our  better data of PG~0946+301 for the existence of
BALs from meta-stable lines.  Our upper limit for $\tau$ C~III$^*$
$\lambda$1175.7 meta-stable line is 0.05. No significant upper limit can be
set for the O~V$^*$ $\lambda$760.4 meta-stable line, since its B
component occurs at the midst of the BAL blend (710--785 \AA) and at
the edge of the G190H data which is of low quality.  Within these
limits no feature is seen in the data. For N~IV$^*$ $\lambda$923
meta-stable line the limit is more tricky to asses. The expected
position of component B for N~IV$^*$ $\lambda$923 (892 \AA) appears at
the edge of an observed absorption feature with $\tau=0.3-0.4$.
However, this absorption feature is very probably due to intervening
absorption systems. Based on observed \Ly\ forest lines (see Appendix~B)
we identify
two intervening O~VI absorption components within this feature, the
red component of one and the blue of the other. The reliability of
this identification is  strengthened by the appearance of the other
doublet component in both systems.  Therefore, it is most likely that
 the observed feature is due to intervening O~VI, but a small
contribution from N~IV$^*$ $\lambda$923 cannot be excluded.  We
conclude that there is no evidence for BALs from meta-stable lines
in the HST data of PG~0946+301. 

It is possible that the absorption feature that Pettini and Boksenberg (1986)
attributed to O~V$^*$ $\lambda$760.4, might be due to absorption from
N~IV $\lambda$765 and/or the S~IV  quadruplet at 750 \AA, which were not
considered in that work.   the IUE data.

\subsection{Micro-physics of the Flow}
A long standing question concerns the existence and properties of a
confining medium for the BAL flow. (Weymann, Turnshek, \& Christiansen 1985;
 Arav \& Li 1994,  Murray et al. 1995, Weymann 1997).
Our current analysis does not shed new light on this question.  The
flow model we described can be made of either  continuous flow components
or have these components made of little cloudlets embedded in a
confining medium. At first glance cloudlets seems to make the low
ionization substrate we invoke unnecessary.  The difference in covering
factor can arise from the difference in sizes between the low
ionization and high ionization region within a given cloudlet.  We
think this is unlikely however,  as such a model  requires
 two forms of
fine tuning. First, the need to have just the right distribution of
cloudlets to make the $\sim2/3$ covering factor for the low ionization
flow (higher cloudlet concentration will make the covering factor
effectively one). Second, a way must be found to avoid different
covering factor for each low ionization line, since there is no
a-priori reason why inside each cloudlets the optically thick zone for
Si~IV and C~III should coincide.  Therefore, it seems more plausible to
create the covering factor of the low ionization flow by the global
geometrical effect we use.

\subsection{The Case for Better Data on PG~0946+301}

As described in the Introduction, the suitability of this
PG~0946$+$301 for IEA studies makes it the prime target for a
multiwavelength campaign. Re-observation of the full rest frame
spectral region (400 -- 1700~\AA\/) using contemporaneous HST/STIS,
FUSE and ground-based spectroscopy will prove crucial to our still
developing understanding of this phenomenon. The 8 ionic transitions
labeled in Figure~5 as well as the important ones of O~IV $\lambda$554
and O~III $\lambda$508 will be available for a much better study
($\S$~3) than we could do using the HST FOS/G160L spectrum.
Saturation limits (see \S~4.1) can be improved by more than an order
of magnitude allowing for a determination of real column densities as
function of velocity.  The spectral coverage of FUSE contain BALs that
can serve as unique saturation diagnostics, enabling us to determine
saturation levels up to a hundred times the apparent optical depth, as
well as BALs from very high ionization states (Si~XII and S~XIV).
X-ray observations would also prove advantageous in helping to
constrain the higher energy ionizing spectrum as well as providing an
estimate of the total line of sight column density.  Variations in the
troughs on time scales of roughly one year have been observed in this
object and so these multiwavelength observations would need to be
contemporaneous.

\section*{ACKNOWLEDGMENTS}
We would like to thank Ross Cohen, Ron Lyons and Tom Barlow
for help in obtaining and reducing the HST and optical observations.)
N. A. and K. K. acknowledge support from STScI grant AR-05784. 
N. A.   acknowledges support from NSF grants 92-23370 and 95-29170.
 M. dK acknowledges support from NSF grant AST 9528256.
M. C. B. acknowledge support from NSF grants AST 95-29175, 91-20599 and NASA
NAGW-3838. Part of this work was performed under the auspices of
the US Department of Energy by Lawrence Livermore National Laboratory
under Contract W-7405-Eng-48.

\renewcommand{\thesubsection}{A.\arabic{subsection}} 

\renewcommand{\theequation}{A.\arabic{equation}}
\setcounter{equation}{0}
\setcounter{subsection}{0}

\section*{Appendix A: Optical depth template extraction}

In this appendix we describe in more detail how we extracted the
optical depth templates for the ions with non-overlapping BAL troughs
that are displayed in figures 2 and 3. Most of these lines (apart from
C III) are multiplets, and the observed troughs are a convolution of the
function that describes the column density as a function of velocity
and the multiplet structure of the line. To probe the structure of the
BAL outflow we want to compare the column densities of the different
ions as a function of velocity, and a deconvolution procedure has to
be applied to the observed BAL troughs. We have experimented with
three different methods described below.

\subsection{Direct inversion method}

The simplest method, which works quite well in most cases, is direct
inversion. After the continuum is determined as described  in
section 4.2,
the spectrum is first normalized to a continuum level of one for all
wavelengths. We then take a region of the spectrum that contains the
BAL trough considered, and extends by at least the width of the
multiplet beyond the region where absorption due to the ion is thought
to be present. This part of the spectrum is then rebinned
to a new wavelength grid that is
equidistant in $\log (\lambda$), and has a bin separation close to the
original one with the constraint that the separation between
the two strongest components of the multiplet corresponds to an integer
number of bins. The logarithmic grid has the advantage that the
separation of the different multiplet components 
remains the same number of bins
as the multiplet is Doppler-shifted. Since our original spectrum is
highly oversampled 
the rebinning does not
degrade the spectrum significantly. 

Now consider the most simple case of a doublet. We first define the
following symbols:
\begin{itemize}
\item  $n$ is the total number of points in the fitted spectrum

\item $s$ is the number of bins corresponding to the doublet
separation

\item  $w_i$\ ($1 \leq i \leq n$) is the wavelength of a
gridpoint on the logarithmic grid

\item $f_i$ is the normalized flux at $w_i$

\item $v_i$ is the velocity that shifts component 1 of the doublet
from rest to the wavelength $w_i$, and component 2 to $w_{i+s}$

\item $ \tau _{k,i}$  ($k=1,2$) is the Sobolev optical depth in 
 component k of the multiplet due to the column density of the ion at velocity
$v_i$:
\begin{equation}
\tau_{k,i} = {{\pi e^2}\over{m_e c}} g_l f_{lu} \lambda {d({{N_l} / {g_l}})\over{dv}},
\end{equation}
where in order to avoid confusion with the algebraic indices 
we use the notation $g_l f_{lu}$ instead of $g_i f_{ik}$.

\item $\tau _i$ is the total optical depth at $w_i$
\end{itemize}

Using these definitions and equation A1, we can now write down
the following equation for $\tau _i$:

$$\tau _i \ = \ \tau _{1,i} \ + \ \tau _{2,i-s} \ = \ \tau _{1,i} +
 \ {{(g_lf_{lu})_2}\over{(g_lf_{lu})_1}} \tau _{1,i-s} \quad $$
as long as we can assume that the population of the lower levels is
proportional to their statistical weights.
Repeating this for all $v_i$ yields a set of coupled linear
equations that we can solve for $ \tau _{1,i}$, for a given vector
$\tau_ i = - \ln (f_i)$. The set of equations can be written
symbolically as  $\rm {\bf A}\ \cdot {\bf \tau}_1 = - \ln ({\bf f})$,
where {\bf A} has non-zero elements only on the main diagonal (equal
to 1) and on the subdiagonal $s$ elements down from the main diagonal
(equal to $(g_lf_{lu})_2/(g_lf_{lu})_1$). 

It is obvious that this method can easily be generalized to multiplets
with an arbitrary number of components. The resulting sets of linear equations
will have a non-zero diagonal for each component of the multiplet, 
filled with the constant value ${{(g_lf_{lu})_k}/(g_lf_{lu})_1}$, and shifted down
or up from the main diagonal by a number of elements equal to the
number of wavelength bins that corresponds to the difference in
wavelength between component 1 and component k. Even for a large
number of points in the spectrum the set of linear  equations can be
solved fast and efficiently by the use of a banded matrix linear
equation solver. 

\subsection{Regularization methods}

Since we are solving for the same number of variables as there are
equations, the direct method forces an exact solution, i.e. the fitted
spectrum is identical to the observed one. This is not always a
desirable property, since for various reasons (very significant noise,
or partial covering effects that influence the doublet ratio) the
observed spectrum does not behave exactly as our theoretical model assumes. 
Requiring an exact match with the given doublet ratio can lead to
instability and oscillating solutions. This effect is not
serious as long as one component of the multiplet has a significantly
higher $g_lf_{lu}$ value than the others, since in this case the oscillations
caused by features in the spectrum that do not follow the expected
line ratio model exactly are damped, and die out quickly.
 
To see this, consider what happens when we try to fit a single strong line
with a doublet model. To get the correct optical depth, the fitting
procedure will put the strongest line of the doublet at the position
of this line. It then expects another absorption line at the position
of the other doublet component, but there is none in the data. To
compensate for the absorption implied by fitting the single line, the
method will then assume negative optical depth in the strongest doublet
component at this wavelength. This in turn predicts emission at the
position of the other doublet line, and an oscillation sets in. When
the $g_lf_{lu}$ value of the primary component is much larger than that of
the secondary, the oscillation will be damped since each
correction is smaller than the previous one by a factor equal to the
ratio of the $g_lf_{lu}$ values. Otherwise, the oscillations will extend over
the entire wavelength range. Thus, this method can cause problems when
extracting column density templates in the case that the multiplet
components have comparable $g_lf_{lu}$ values.

However, since there are physically reasonable models that could lead
to equal ``effective'' optical depth in each component of the
multiplet, such as the case in which the line depth is caused by very
optically thick clouds that cover the source only partly, it is
interesting to extract templates under this assumptions. We have
experimented with two methods to stabilize the solution.

The first is the linear regularization technique described in Press et
al. (1992). For details we refer to this reference, but the basic idea
is that one does not minimize the difference between the fit and the
model alone (i.e. solve the equation  $\rm {\bf A}\ \cdot {\bf \tau}_1
= - \ln {\bf f}$), but rather attempts to find the best fitting
solution while also getting as close as possible to some 
{\it a priori} constraint, for instance that the solution should be
relatively smooth. As derived in Press et al., this leads to a set of
linear equations of the form
\begin{equation}
 \rm ({\bf A}^T\ \cdot {\bf A}\ + \ \lambda {\bf H})\cdot
  {\bf \tau}_1 = - {\bf A}^T \cdot \ln {\bf f}
\end{equation}
in which the matrix {\bf H} describes the {\it a priori} constraint,
and the parameter $\lambda$ allows a trade-off between fitting as
close as possible to the original equation ($\lambda=0$) or to the
constraint ($\lambda>>1$). We experimented with the form of {\bf H},
and found good regularization for the choice $h_{i,i}=4$, $h_{i-1,i}=-1$,
$h_{i-2,i}=-1$, $h_{i+1,i}=-1$, $h_{i+2,i}=-1$ and all other elements
equal to zero. For $\lambda$ we used the value 0.5.

The second method we tried was modeling the $\tau_{1,i}$ distribution
by a large number of gaussians with a width comparable to the
resolution, i.e. about 4 datapoints. Each gaussian in the primary
component of the multiplet implies similar gaussians at the
wavelengths corresponding to the other multiplet lines, the amplitudes
again being proportional to the $g_lf_{lu}$ values. We then used a least
squares method based on singular value decomposition to obtain the 
optical depths of the primary multiplet component that best fit the
spectrum. 

In figure 11 we plot the results of an extraction of the column
density profile for the C IV line under the assumption that the
doublet ratio is 1 instead of 0.5 . Clearly the direct method is
unstable, and unusable for this profile. The method of Gaussian fits
damps out a lot of the instability, but still shows a considerable
amount of oscillation at the doublet separation. The regularization
method performs best, yielding a column density distribution
without negative excursions, and relatively little oscillation. It
should be noted however that the regularization does involve smoothing
and loss off information. 

 \begin{figure}
\centerline{\psfig{file=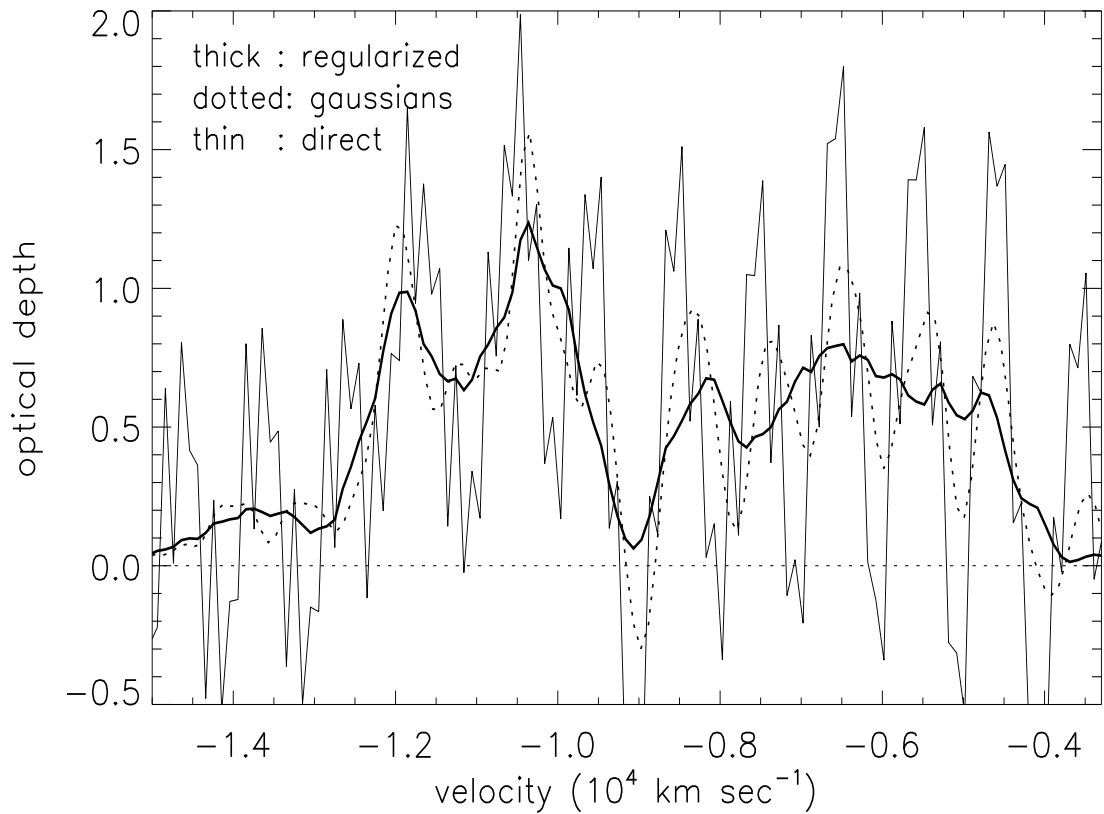,height=12.0cm,width=16cm}}

\caption{The results of applying three different
deconvolution methods to the observed line profile of C~IV, under the
assumption that the effective doublet ratio is 1. Both direct
inversion and multiple Gaussian fitting introduce oscillations, only
the regularization method gives physically acceptable results. }
\label{fig A1}  
\end{figure}

The three methods were also tested by applying them to simulated line
profiles with added noise or constructed with the wrong line ratios. 
The direct method was found to perform best since it does not manipulate the
original data. However,  it is most likely to introduce
oscillations if the line ratios are wrong. The
regularization is the best for avoiding the instability associated with
equal strength components and noise. We found that the gaussian
fitting method is good at suppressing the noise while still showing
oscillations if the line ratios are incorrect, which is a useful
diagnostic tool.

\subsection{Noise and spurious features} 

 \begin{figure}
\centerline{\psfig{file=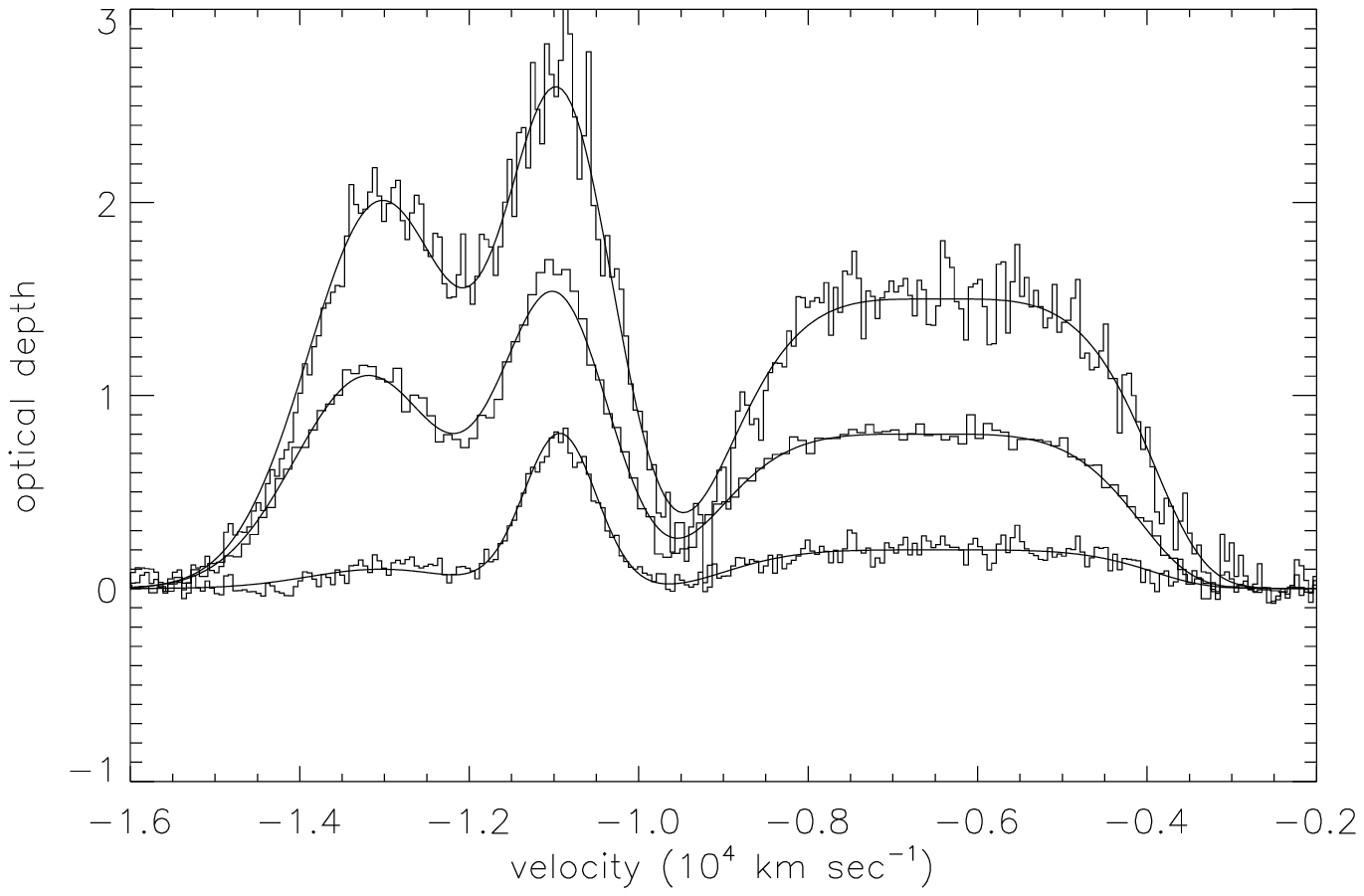,height=12.0cm,width=16cm}}

\caption{ This figure illustrates the expected level of fluctuations in the
deconvolved profiles of optical depth as a function velocity due to
noise in the observations. A smooth underlying optical depth profile 
(the thin line in the figure) was used to generate an absorption line
profile, which was then degraded to a signal to noise level corresponding to
the ones for O~VI, C~IV and Si~IV. The resulting line profile was then
deconvolved using direct inversion, just as it was done for the real
profiles in Figure \ref{fig-2}. Features in the real profile
comparable or smaller than those present in the simulation shown here
are not significant. }\label{fig A2}  
\end{figure}

To estimate what features in the extracted templates are real and
which ones are likely to be due to noise, we extracted templates from
artificial spectra that were realizations (with a realistic signal to
noise ratio) of a smooth spectrum similar in shape to the observed
one.  The results for O VI, C IV and Si IV are plotted in figure 12,
which can be compared with the real spectra in Figure \ref{fig-2}.
For these lines, the S/N ratios in the continuum are about 18, 45 and
24 respectively.  From this comparison e conclude that in the broad
component the O VI profile does not contain measurable substructure,
but that the substructures in the C~IV and Si IV templates are
significant.

Even if structures are significant, they need not necessarily be reflections
of features in the column density distribution, since there are also
other effects apart from noise that could influence the templates, such
as lines of other species, or components that are optically thick but
partially covering the source and thus do not fit the assumed 
multiplet structure. A good example of this is the C IV template
for a doublet ratio of one obtained by the Gaussian fitting method 
in figure 11. However, moderate departures from the theoretical
line ratios (say 0.7 instead of 0.5) do not lead to large apparent
substructures.

\renewcommand{\thesubsection}{B.\arabic{subsection}} 
\setcounter{subsection}{0}

\section*{APPENDIX B: INTERVENING ABSORPTION}

 As in all quasars, absorption features from intervening material are
also seen in the spectrum of PG 0946+301.  In this appendix we attempt
to catalogue the intervening absorption features in order to avoid
confusion with BAL features, facilitate future studies of this
object, and  account for all the absorption features seen in the spectrum.

\subsection{Ly$\alpha$  systems}
We identify 14 suspected \Ly\ absorption systems. Three of these (5, 9
and 10) are clearly seen in other lines and two more might be seen in
Ly$\beta$ (7 and 12).  Observed wavelength and equivalent widths for
the Ly$\alpha$ systems are given in Table 3. We note that system 1
might be a BAL feature associated with component B of Si~III
$\lambda$1206.5 but we classify it as an intervening \Ly\ absorption
since it is significantly narrower than other BAL features associated
with component B.  Also, \Ly\ lines 4 and 8 are quite marginal with the
available S/N.

In Table 3 we also give a subjective assessment for the reality of
absorption features that can arise from other lines associated with
each intervening Ly$\alpha$ line.  It is interesting to notice how
much of the non-BAL absorption features around 1920--1980 \AA\ can be
explained by lines associated with the Ly$\alpha$ systems (see
figure 13).  In particularly system 10 has a strong Ly$\beta$ and O~VI
lines in that region (and also a clear C~IV doublet).  The most
intriguing is system 9, which is too broad too be a simple Ly$\alpha$
forest line.  Some of the absorption features that can be explained by
system 10 around 1960--1990 \AA\ can also arise from system 9.
Furthermore, the absorption feature at 840 \AA\ (rest frame, see
Fig. 5) is consistent with being the Ly$\gamma$ of system 9.
However, system 9 doesn't have a definite C~IV component, although it
might be somewhat obscured by the edge of the Si~IV BAL. System 5 is
clearly seen in C~III, C~IV, and may be seen in O~VI. C~IV absorption
associated with system 10 appears at the high velocity edge of the
Si~IV BAL and should not be confused as being a part of it. We do not find
 N~V or Si~IV lines associated with any of the Ly$\alpha$ systems.
 \begin{figure}
\centerline{\psfig{file=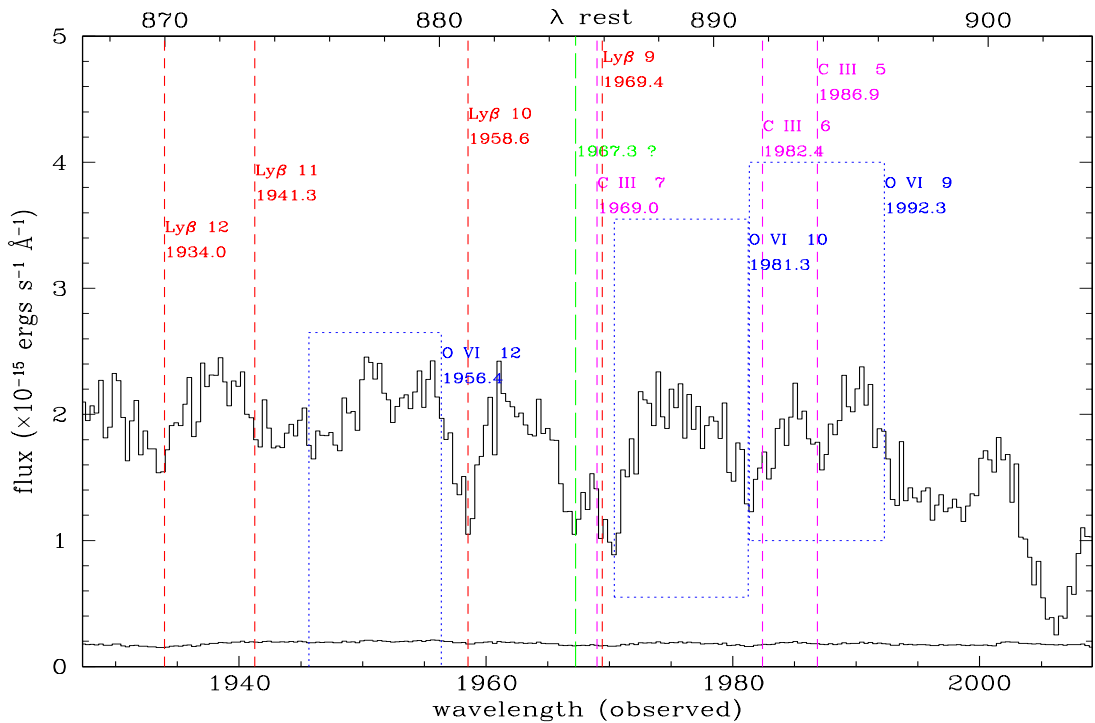,angle=-90,height=10.0cm,width=16cm}}
\caption{ Segment of PG 0946+301 spectrum with intervening absorption
marked.  Vertical lines designate the calculated positions of metal
lines and Ly$\beta$, which correspond to  identified \Ly\ systems
(see table 3).  Ly$\beta$ and C~III lines are marked with dashed
lines. The ion and the number of the system is written just to the
right of the line and its calculated wavelength is written underneath.
The O~VI doublets are marked with a rectangular dotted box where the
vertical sides correspond to the positions of the doublet
components. The label of the system and the calculated wavelength of
the red component are attached to the right of the component. To avoid
cluttering we omitted most of the non detected lines. About 75\% of
the absorption features seen in this part of the spectrum can be
identified with lines associated with known \Ly\ intervening systems.
Absorption seen longward of 1990 \AA\ is part of the S~VI BAL.  }\label{fig-apb}
\end{figure}
 
\subsection{Galactic Lines}
A strong Mg II doublet ($\lambda \lambda$2796.35, 2803.32)
 is seen in the observed frame and we used it to calibrate 
the the observed wavelengths.
Two systems of galactic Fe II absorption lines at almost zero redshift
are seen in the spectrum.  Working in the frame where the observed
galactic Mg II is at zero velocity, we find that observed absorption features 
 are matching well with Fe II lines (2344.21, 2374.46 and 2382.76 \AA) 
The second  identified system is the two Fe~II zero
energy level transitions at 2586.65 and 2600.17 \AA. 
The absence of other
components of the Fe II multiplets  is explained by noticing
that all these components are not truly ground state transitions, but arise
from slightly higher energy states.  

\subsection{Unidentified Lines}

Very few absorption features in the spectrum are left unexplained once
the BAL absorption and the different intervening systems are accounted
for.  The feature at 1967.3 \AA\ is probably another Ly$\alpha$ forest
line.  The absorption features seen at 2138 \AA\ and 2177.4 \AA\ are
part of the BAL flow.  They match  component E of the flow in
C~III and N~III respectively.  We identified this component (the
lowest velocity one) in all 7 BALs for which we have an optical depth
solution.  The very broad feature at 2305.9 \AA\ should be a BAL
feature.  There is a very good agreement with the expected position of
a S~IV BAL, but this is a relatively weak line and its stronger
brothers to the blue should be much deeper.  As noted in \S~5.7, this
is one of the most puzzling issues in the BAL features of this object

\begin{table}
\begin{center}
\begin{tabular}{ccccccccc}
\multicolumn{9}{c}{\sc Table 3: Intervening lines} 
\\[0.2cm]
\hline
\hline
%
\multicolumn{1}{c}{no.}
&\multicolumn{1}{c}{$\lambda$(\Ly)}
&\multicolumn{1}{c}{EW(\Ly)}
&\multicolumn{1}{c}{Ly$\beta$}
&\multicolumn{1}{c}{C III}
&\multicolumn{1}{c}{C IV$_b$}
&\multicolumn{1}{c}{C IV$_r$}
&\multicolumn{1}{c}{O VI$_b$}
&\multicolumn{1}{c}{O VI$_r$}
\\
\multicolumn{1}{c}{} & \multicolumn{1}{c}{(obs)} & 
\multicolumn{1}{c}{(rest)} & \multicolumn{1}{c}{} &
\multicolumn{1}{c}{} & \multicolumn{1}{c}{} & 
\multicolumn{1}{c}{} & \multicolumn{1}{c}{} & 
\multicolumn{1}{c}{} 
\\[0.05cm]
\hline
1 &  2592.7 &  0.20  &     0  &     0  &     1b &     0b  &    0b   &   0b  \\
2 &  2517.5 &  0.24  &     1b &     0  &     0  &     0   &    1b   &   0  \\
3 &  2514.9 &  0.23  &     1b &     0  &     0  &     0   &    0    &   0  \\
4 &  2494.9 &  0.10  &     1  &     0b &     1  &     1   &    0    &   0  \\   
5 &  2472.2 &  0.48  &     1  & {\bf3} & {\bf3} & {\bf3}  &    2    &   1b  \\ 
6 &  2466.6 &  0.37  &     1  &     2b &     1  &     1   &    0    &   1  \\  
7 &  2450.0 &  0.32  &     2b &     0b &     0  &     1   &    1    &   2  \\ 
8 &  2351.9 &  0.15  &     0  &     0  &     0  &     0b &     0b &     0b  \\  
9 &  2334.2 &  1.65  &     2b &     0  &     1  &     1  & {\bf3} &     2b  \\  
10 & 2321.3 &  0.82  & {\bf3} &     1  & {\bf3} & {\bf3} & {\bf3} & {\bf3}  \\   
11 & 2300.8 &  0.14  &     1  &     0  &     0  &     0  &     1  &     1  \\     
12 & 2292.1 &  0.52  &     2  &     0  &     0  &     0  &     1    &   1  \\  
13 & 2283.9 &  0.29  &     0  &     0  &     0  &     0  &     0    &   0  \\  
14 & 2153.2 &  0.44  &     1  &     0  &     0  &     0  &     1    &   2b  \\ 
\\[0.01cm]
\hline
\end{tabular}
\end{center}
 Column 1
gives the number the system, column 2 the observed
wavelength of the \Ly\  line and column 3 gives its rest
equivalent width. In  columns 4--9 we give a qualitative
assessment for the existence of an absorption feature in other lines
from the same system.  Our key is as follows: 0 - no feature, 1 -
marginal, 2 - an absorption feature exists but it either not strong
enough for an unambiguous identification or its position does not
precisely match the expected one, 3 unambiguous detection at the
expected position and with a similar profile. The letter b designate a
feature blended with other absorption (mostly BAL but not always). For
C~IV and O~VI the subscript designate the red or blue component of the
doublet.
\end{table}

%
%

\end{document}